\documentclass[twocolumn,journal]{IEEEtran}
\usepackage[T1]{fontenc}
\usepackage[latin9]{inputenc}
\usepackage{color}
\usepackage{float}
\usepackage{amsmath}
\usepackage{amsthm}
\usepackage{amssymb}
\usepackage{mathdots}
\usepackage{graphicx}
\usepackage[unicode=true]
 {hyperref}

\makeatletter

\floatstyle{ruled}
\newfloat{algorithm}{tbp}{loa}
\providecommand{\algorithmname}{Algorithm}
\floatname{algorithm}{\protect\algorithmname}

 \let\oldforeign@language\foreign@language
 \DeclareRobustCommand{\foreign@language}[1]{%
   \lowercase{\oldforeign@language{#1}}}
\theoremstyle{plain}
\newtheorem{thm}{\protect\theoremname}
\theoremstyle{definition}
\newtheorem{defn}[thm]{\protect\definitionname}
\theoremstyle{definition}
\newtheorem{problem}[thm]{\protect\problemname}
\theoremstyle{remark}
\newtheorem{rem}[thm]{\protect\remarkname}
\theoremstyle{plain}
\newtheorem{lem}[thm]{\protect\lemmaname}
\theoremstyle{plain}
\newtheorem{cor}[thm]{\protect\corollaryname}

\usepackage[caption=false,font=footnotesize]{subfig}
\usepackage{algorithmic}
\usepackage{algorithm}

\allowdisplaybreaks

\makeatother

\providecommand{\corollaryname}{Corollary}
\providecommand{\definitionname}{Definition}
\providecommand{\lemmaname}{Lemma}
\providecommand{\problemname}{Problem}
\providecommand{\remarkname}{Remark}
\providecommand{\theoremname}{Theorem}

\begin{document}

\title{Low-Overhead Hierarchically-Sparse Channel Estimation
for Multiuser Wideband Massive MIMO}

\author{Gerhard Wunder, Stelios Stefanatos, Axel Flinth, Ingo Roth, and Giuseppe Caire 
	
\thanks{ GW and SS acknowledge support from H2020 project ONE5G (ICT-760809)
receiving funds from the European Union. The authors would like to
acknowledge the contributions of their colleagues in the project,
although the views expressed in this contribution are those of the
authors and do not necessarily represent the project. AF acknowledges support from the DFG (Grant KU 1446/18-1) and ANR
JCJC OMS, IR from the DFG (EI 519/9-1), the Templeton Foundation and the
ERC (TAQ),  and GW from the DFG (WU 598/7-1 and WU
598/8-1). All DFG projects are within the German priority program
on ``Compressed Sensing in Information Processing'' (COSIP). }

\thanks{G. Wunder and S. Stefanatos are with the Department of Mathematics and Computer Science, Freie Universit{\"a}t Berlin, Berlin, Germany (email: g.wunder@fu-berlin.de,  stelios.stefanatos@fu-berlin.de).}

\thanks{A. Flinth is with the Institut de Math\'{e}matiques, Universit\'{e} de Toulouse III Paul Sabatier, Toulouse, France}

\thanks{I. Roth is with the Dahlem Center for Complex Quantum Systems, Freie Universit{\"a}t Berlin, Germany.}

\thanks{G. Caire is with the Department of Electrical Engineering and Computer Science, Technical University of Berlin, 10623 Berlin, Germany (email: caire@tu-berlin.de).}

\thanks{A preliminary version of some of the results reported in this work appeared in \cite{WSA}.}

}

\markboth{}{}
\maketitle
\begin{abstract}
Numerical evidence suggests that compressive sensing (CS) approaches for wideband massive MIMO channel estimation can achieve very good performance with limited training overhead by exploiting the sparsity of the physical channel. However, analytical characterization of the (minimum) training overhead requirements is still an open issue. By observing that the wideband massive MIMO channel can be represented by a vector that is not simply sparse but has well defined structural properties, referred to as hierarchical sparsity, we propose low complexity channel estimators for the uplink multiuser scenario that take this property into account. By employing the framework of the hierarchical restricted isometry property, rigorous performance guarantees for these algorithms are provided suggesting concrete design goals for the user pilot sequences. For a specific design, we analytically characterize the scaling of the required pilot overhead with increasing number of antennas and bandwidth, revealing that, as long as the number of antennas is sufficiently large, it is independent of the per user channel sparsity level as well as the number of active users. These analytical insights are verified by simulations demonstrating also the superiority of the proposed algorithm over conventional CS algorithms that ignore the hierarchical sparsity property.
\end{abstract}

\begin{IEEEkeywords}
massive MIMO, OFDM, channel estimation, compressed/compressive sensing,
training overhead, multiuser, hierarchical sparsity
\end{IEEEkeywords}

\IEEEpeerreviewmaketitle{}

\section{Introduction }

\IEEEPARstart{M}{assive} mutliple-input multiple-output (MIMO) is
the term used to describe the practice of deploying a large number
of antennas at the base station (BS), which is considered as a key
technology for 5G \cite{5g2017_JSAC}. Although the benefits of massive
MIMO are by now well understood \cite{mMIMO book}, the fundamental
bottleneck for massive MIMO deployment in a multi-cell scenario is
pilot contamination, i.e., degradation of the uplink channel state
information (CSI) due to multiple user equipments (UEs) transmitting
non-orthogonal training signals on the same set of resources \cite{Marzetta mMIMO}.
In addition, with the emergence of massive machine type communications
(MTCs) with typically small data bursts, there is a need to decrease
the signaling overhead associated with the CSI acquisition \cite{mMTC mag},
thus resulting in pilot contamination issues also in a single-cell
scenario. It is therefore of critical importance to come up with designs
that balance the conflicting requirements of accurate  CSI and
low training overhead, for systems with a massive number of antennas,
UEs, and bandwidth.

\subsection{Related Work}

The topic of (optimal) training design and channel estimation for
the single UE case has been extensively studied for the multi-antenna
and/or wideband (OFDM) channels both from an estimation mean squared
error (MSE) as well as a capacity perspective (see, e.g., \cite{Giannakis optimal training OFDM,Leus optimal training MIMO OFDM,Viswanathan optimal training,Hassibi how much training}).
This line of works on pilot-aided system design was based on the assumption
of a rich scattering propagation environment, {effectively treating the channel response among different antennas as independent}. This results in pilot
designs having a training overhead that is proportional to the product
of the number of antenna elements and the system bandwidth. Application of
these approaches in the multiuser setting may be unacceptable due
to limited resources that cannot allow for orthogonal pilot transmissions,
resulting in pilot contamination effects \cite{Marzetta mMIMO,pilot contamination survey paper}.

The key towards reducing the training overhead is the observation
that the wireless channel is fundamentally \emph{sparse}, i.e., a
signal arrives at the receiver via a limited number of distinct (resolvable)
paths \cite{Sayeed MIMO channel model}. This propagation has been experimentally observed to hold true with large carrier frequencies  (beyond 2GHz) and/or with large antenna array (i.e., massive MIMO) \cite{cost2100}. 
 Therefore, posing the channel
estimation problem as that of identifying the channel paths properties
(gain, delay, angle) immediately implies improvement of the CSI procedure
over the conventional approaches, either in terms of performance (MSE)
or training overhead, as the number of unknowns to be estimated (significantly)
decreases.

Earlier works exploiting this channel sparsity for estimation purposes
(e.g., \cite{Letaief parametric channel est}) utilized traditional
tools from the fields of array processing and harmonic retrieval \cite{Stoica book},
however, the focus was only on algorithmic and performance aspects,
and the issue of training overhead minimization was ignored. The recent
advent of the field of compressive sensing (CS) \cite{CS signal process. mag},
which considers the problem of solving an under-determined linear system
under the assumption that the vector to be estimated is sparse, has
provided a new set of tools towards low-overhead sparse channel estimation
\cite{CS for channel est Mag}. Considering the problem of wideband
massive MIMO channel estimation, by reformulating it in a format compatible
to the one considered in CS, a few recent publications have proposed
CS-inspired channel estimation algorithms demonstrating that excellent
performance is indeed possible with low training overhead \cite{Swindlehurst2016_TSP,Heath JSAC 2017}.
However, these performance results are only provided via means of
numerical simulations, with very limited (if at all) analytical insights
on the \emph{training overhead required to achieve a certain performance}. A different approach is considered in \cite{Haghighatshoar2017_TWC}, where it is the (sparse) covariance matrix of the channel that is estimated by a CS approach, with the resulting estimate used to perform linear minimum mean squared error (LMMSE) channel estimation. However, this approach requires the observation of multiple, independent channel snapshots, which might not be possible under certain scenarios (e.g., short length MTCs).

Intuitively, one expects that the utilization of multiple antennas
at the BS can aid in reducing the bandwidth dedicated for training
signals. However, to the best of our knowledge, there are no analytical
results available that confirm this intuition, even though CS theory
provides numerous rigorous answers regarding number of measurements
required to achieve good estimation performance \cite{MathIntroToCS}.
{This lack of analysis in the considered setup is mainly due to the,
so called, sensing matrix of the corresponding CS problem formulation
having a Kronecker-product structure \cite{Heath how many measurements}. Even though Kronecker-product sensing matrices have been explicitly investigated in the CS literature \cite{Jokar Kronecker, Duarte Kronecker}, the available results suggest a required training overhead that is overly pessimistic (cf. discussion of Theorem 9).}

\subsection{Contributions}

In this paper, a setup with a single BS equipped with a uniform linear
array (ULA) serving multiple single-antenna uplink UEs is considered,
with the goal of proposing efficient channel estimation algorithms
as well as providing rigorous analytical insights on the overhead
requirements. {Under the assumption of, so called, on-grid channel
parameters, that is reasonable for asymptotically large number of antennas
and bandwidth, a key observation is that the channel estimation problem
can be formulated as the CS estimation of a vector that is not simply
sparse but \emph{hierarchically sparse} \cite{C-HiLasso, Bockelmann,  Roth2016_TSP,RothEtAl2018_ISIT}. In particular, the positions of its non-zero elements cannot be arbitrary but are subject to constraints implied by the physical channel properties. This is a critical property that is exploited in the following.}
%but whose support has a well defined structure, referred to
%as\emph{ hierarchical sparsity}, that is directly suggested by the
%physical channel properties. 

The main contributions of the paper are
summarized as follows. 
\begin{itemize}
\item Two novel, low-complexity channel estimation algorithms are proposed
that explicitly take into account the hierarchical sparsity property,
which was ignored in the previous literature. The algorithm description
is provided for an arbitrary pilot sequence design, where multiple
UEs utilize the same subcarriers of a single OFDM symbol for training
purposes.
\item The notion of the hierarchical restricted isometry property (HiRIP) is
introduced, which can be considered as a specialization of the standard
RIP notion \cite{MathIntroToCS} to the setting of hierarchically sparse vectors. Rigorous
guarantees for reliable, i.e., bounded error, channel estimation by
the proposed algorithms are provided based on the, so called, HiRIP
constant of the Kronecker-product type sensing matrix of the corresponding
CS estimation problem. 
\item The above characterization provides a concrete design goal for the
pilot sequence design, namely, it should be such that the HiRIP constant of
the sensing matrix is sufficiently small. Towards this, a design based
on phase-shifted UE pilot sequences is proposed, which allows for
a rigorous description of the scaling of the number of pilot subcarriers
and number of observed antennas required to achieve reliable channel
estimation. The analysis highlights the benefit of using multiple
antennas in the sense of allowing for reduced pilot-overhead compared
to the single antenna case. Even more important, for sufficiently
large number of antennas, the pilot overhead required is independent
of the number of (active) UEs and number of channel paths per UE.
These conclusions are verified by numerical simulations demonstrating
also the superior performance of the proposed algorithms compared
to standard CS algorithms of comparable complexity that ignore hierarchical
sparsity. The latter requires a significantly larger minimum pilot
overhead to achieve reasonable performance  that also increases with
number of channel paths per UE.
\item The cases of jointly processing multiple training OFDM symbols as
well as channels with off-grid parameters is also discussed, as both
can be naturally accommodated by the proposed framework. Simulations
show that in the latter case, although the mismatch of assuming on-grid
channel parameters by the algorithm results in a performance degradation,
performance remains still significantly better in terms of required
overhead compared to standard CS algorithms of comparable complexity
as well as the conventional linear minimum mean square error (LMMSE)
estimator that ignores channel sparsity altogether.
\end{itemize}

\subsection{Notation}

Vectors and matrices will be denoted by lower and upper case bold
letters, respectively. All vectors are column vectors. The $(n,m)$
element of $\mathbf{X}\in\mathbb{C}^{N\times M}$ is denoted by $[\mathbf{X}]_{n,m},n\in[N],m\in[M],$
with $[N]\triangleq\{0,1,\ldots,N-1\}$. $(\cdot)^{*},(\cdot)^{T},(\cdot)^{H}$
denote complex conjugate, transpose, and Hermitian operation, respectively.
$\lVert\mathbf{X}\rVert\triangleq\sqrt{\text{tr}\{\mathbf{X}^{H}\mathbf{X}\}}$
is the Frobenius norm (Euclidean norm if $\mathbf{X}$ is a vector).
The cardinality of a set $\mathcal{A}$ is denoted by $|\mathcal{A}|$.
$\mathbf{X}_{\mathcal{A}}$ ($\mathbf{x}_{\mathcal{A}}$) denotes
the matrix (vector) obtained either by extracting the rows (elements)
of $\mathbf{X}$ ($\mathbf{x}$) enumerated by $\mathcal{A}\subseteq[N]$
or by setting the rows (elements) of $\mathbf{X}$ ($\mathbf{x}$)
that do not belong to $\mathcal{A}$ equal to zero (the case will
be clear from the context). The $N\times N$ identity matrix is denoted
by $\mathbf{I}_{N}$ and $\text{diag}(\mathbf{x})$ denotes the diagonal
matrix with $\mathbf{x}$ on its diagonal. $\mathbf{F}_{N,M}$ denotes
the matrix obtained by the first $M\leq N$ columns of the $N\times N$
DFT matrix, i.e., $[\mathbf{F}_{N,M}]{}_{n,m}\triangleq e^{-j2\pi mn/N},n\in[N],m\in[M]$.
The vector resulting of stacking the columns of a matrix $\mathbf{X}$
is denoted by $\text{vec}(\mathbf{X})$. $\text{supp}(\mathbf{x})\subseteq[N]$
denotes the set of non-zero elements (support) of $\mathbf{x}\in\mathbb{C}^{N}$.
$\mathbb{C}^{N_{1}\cdot N_{2}\cdots N_{\ell}}$ denotes the space
of complex-valued, multilevel block vectors consisting of $N_{1}$
blocks, each containing $N_{2}$ blocks, $\ldots$, each containing
$N_{\ell-1}$ blocks of $N_{\ell}$ elements (for a total of $N_{1}N_{2}\cdots N_{\ell}$
elements). A vector $\mathbf{x}$ is called $s$-sparse if $|\text{supp}(\mathbf{x})|=s$.\textcolor{red}{{}
}For reference, the following standard definition from CS theory \cite{MathIntroToCS}
is recalled below.
\begin{defn}[RIP constant]
\label{def: standard RIP definition}The restricted isometry constant
$\delta_{s}(\mathbf{A})$ of a (deterministic) matrix $\mathbf{A}\in\mathbb{C}^{N\times M}$
is the smallest $\delta\geq0$ such that 
\begin{equation}
(1-\delta)\Vert\mathbf{x}\Vert^{2}\leq\Vert\mathbf{A}\mathbf{x}\Vert^{2}\leq(1+\delta)\Vert\mathbf{x}\Vert^{2},\label{eq:standard RIP definition}
\end{equation}
for all $s$-sparse vectors $\mathbf{x}\in\mathbb{C}^{M}$ ($s\leq M$).
We say that $\mathbf{A}$ satisfies the ($s$-th) restricted isometry
property ($s$-RIP) if $\delta_{s}(\mathbf{A})<\bar{\delta}$ where $\bar{\delta}<1$ is a pre-specified constant.
\end{defn}

\section{Wideband \textcolor{black}{Massive MIMO }Channel Model and Delay-Angular
Representation}

We consider the uplink of a single cell with a BS equipped with $M\gg1$
antenna elements serving multiple single-antenna UEs.
For a ULA, the array manifold $\mathbf{a}\left(\cdot\right):\left[-\pi/2,\pi/2\right]\rightarrow\mathbb{C}^{M}$,
which maps angular to spatial domain, is given by $\mathbf{a}\left(\phi\right)\triangleq[1,e^{-j2\pi d\sin\phi},\ldots,e^{-j2\pi d\left(M-1\right)\sin\phi}]^{T}$
\cite{ChenYang206_TWC}. Here, $d$ is the normalized spatial separation
of the ULA (with respect to carrier wavelength), which, without loss
of generality (w.l.o.g.), is assumed to be equal to $1/2$ in the following.
As is routinely done, we perform the change of variable $\theta=d\sin(\phi)\in[-1/2,1/2]$
and, with a slight abuse of notation, we write the array manifold
as a function of $\theta$, i.e., $\mathbf{a}\left(\theta\right)=[1,e^{-j2\pi\theta},\ldots,e^{-j2\pi\left(M-1\right)\theta}]^{T}$.
Noting that $\mathbf{a}(\theta)=\mathbf{a}(1-\theta)$ for $\theta<0$,
it is convenient to treat $\theta$ as taking values in $[0,1]$.
Considering a sampled version of this interval by the $M$ points
$\{k/M\}_{k=0}^{M-1}$ yields the steering (dictionary) matrix $\mathbf{A}_{\theta}\triangleq\left[\mathbf{a}(0),\mathbf{a}(1/M),\ldots,\mathbf{a}((M-1)/M)\right]=\mathbf{F}_{M,M}\in\mathbb{C}^{M\times M}$. 

Transmissions are performed via wideband OFDM signals with $N\gg1$
subcarriers centered at the \textcolor{black}{baseband} frequencies
$\{2\pi k/T_{s}\}_{k=0}^{N-1}$, with $T_{s}>0$ being the \textcolor{black}{useful
(without the cyclic prefix) }OFDM symbol duration. Assuming that the
maximum delay spread of \textcolor{black}{all UE} channels is not longer 
than $\alpha T_{s},\alpha\leq1$, which is the case in any reasonable
OFDM design, the delay manifold $\mathbf{b}\left(\cdot\right):\left[0,\alpha T_{s}\right]\rightarrow\mathbb{C}^{N}$,
which maps the delay to the frequency domain, is defined as $\mathbf{b}\left(\tau\right)\triangleq[1,e^{-j2\pi\tau/T_{s}},\ldots,e^{-j2\pi(N-1)\tau/T_{s}}]^{T}$
\cite{ChenYang206_TWC}. Considering a\textcolor{black}{{} sampled version}
of $[0,T_{s}]$ by the $N$ points $\{kT_{s}/N\}_{k=0}^{N-1}$, yields
the steering (dictionary) matrix $\mathbf{A}_{\tau}\triangleq\left[\mathbf{b}(0),\mathbf{b}(T_{s}/N),\ldots,\mathbf{b}((D-1)T_{s}/N)\right]=\mathbf{F}_{N,D}\in\mathbb{C}^{N\times D}$
where $D\triangleq\lfloor\alpha N\rfloor$ is the channel delay spread
in samples.\footnote{In general, a denser sampling for the angle and delay domains could
be employed. We leave investigations of this case to future work.}

The channel of an arbitrary UE is a superposition
of a small number $L$ of impinging wavefronts (paths) characterized
by their delay/angle pairs $\{(\tau_{p},\theta_{p})\}_{p=0}^{L-1}$,
with $\tau_{p}\in[0,\alpha T_{s}]$, $\theta_{p}\in[0,1]$. This is
reflected in the channel transfer matrix $\mathbf{H}\in\mathbb{C}^{N\times M}$
whose $(n,m)$-th element corresponds to the complex channel gain at
subcarrier $n$ and antenna $m$ and can be written as \cite{Haghighatshoar2017_TWC,ChenYang206_TWC}
\begin{equation}
\mathbf{H}=\sum_{p=0}^{L-1}\rho_{p}\mathbf{b}\left(\tau_{p}\right)\mathbf{a}^{H}\left(\theta_{p}\right),\label{eq:channelmatrix}
\end{equation}
where $\rho_{p}\in\mathbb{C}$ is the complex gain of the $p$-th
path. It is noted that $L$ is treated here as a given  parameter that depends only on the physical propagation properties and is independent of system parameters $M$ and $N$.

Targeting low-complexity channel estimation, \textcolor{black}{it
is beneficial to} consider an \textcolor{black}{alternative} representation
of $\mathbf{H}$, which translates the physical sparsity to sparsity
of an appropriately defined matrix that is to be identified by the
estimator. Towards this end, we will first \textcolor{black}{consider
the case of }\textcolor{black}{\emph{on-grid}}\textcolor{black}{{} channel
parameters,} when every delay/angle pair lies exactly on the delay/angle
grid corresponding to the steering matrices $\mathbf{A}_{\theta}$
and $\mathbf{A}_{\tau}$, i.e., it holds $(\tau_{p},\theta_{p})=(k_{p}T_{s}/N,l_{p}/M)$
for some $k_{p}\in[D]$ and $l_{p}\in[M]$, for all $p\in[L]$. In
general, this assumption does not hold, however, it is a reasonable
approximation for asymptotically large $N$, $M$, and is convenient
for algorithm design and (asymptotic) performance analysis. The more
general, \emph{off-grid} channel parameters case will be treated in
Sec. V. Note that, apart from the on-grid/off-grid  delay/angle pairs characterization, no assumptions on the (joint) statistics of path delays, angles, and gains are considered in the following treatment.  

\begin{figure}[ptb]
\noindent \centering{}\includegraphics[width=1\columnwidth]{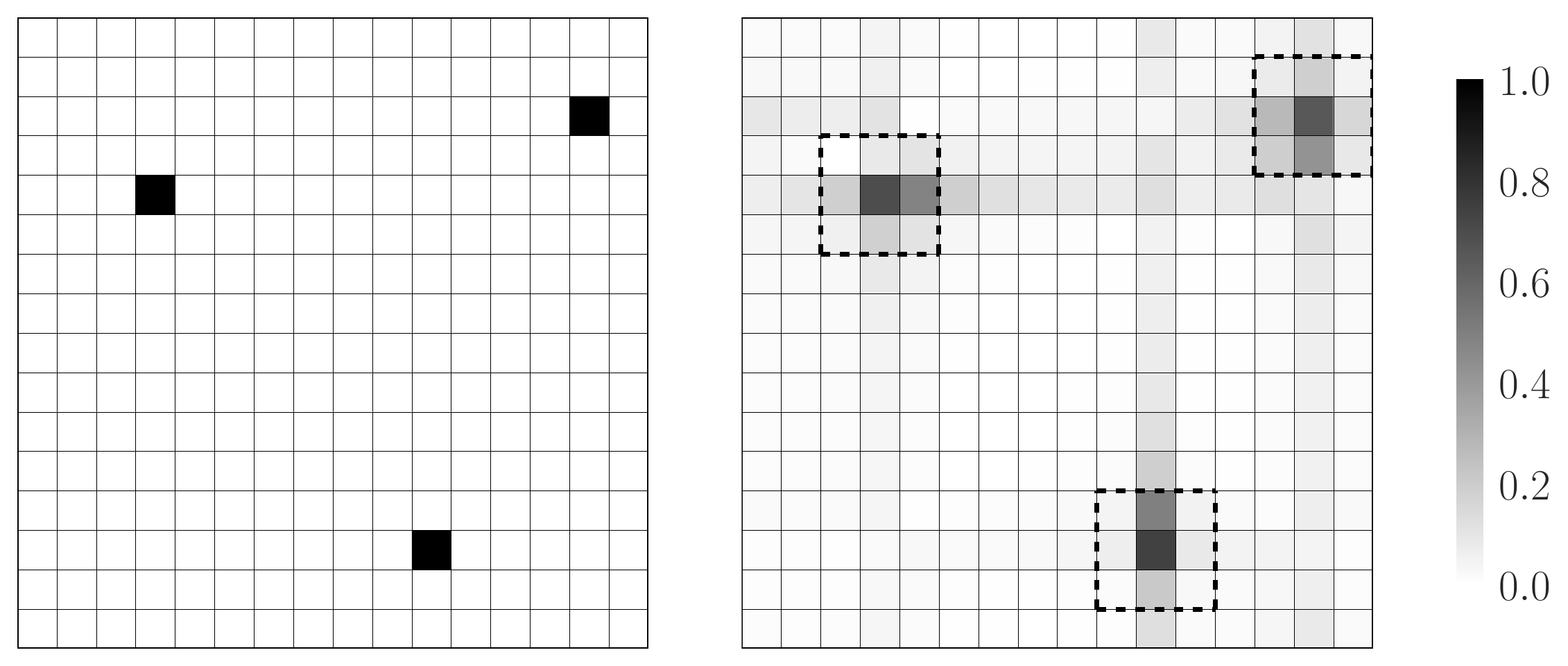}\caption{Example heatmap (modulus values) for the delay-angular representation
$\mathbf{X}$ of a channel with $L=3$ and $\rho_{p}=1$ for all $p\in[L]$,
and $N=M=D=16$. Left: on grid case; Right: off-grid case, obtained
by slight perturbation of the angle/delay pairs values of the on-grid
case.}
\label{fig grid and off-grid W}
\end{figure}

With on-grid parameters,  $\mathbf{H}$ can
then be written as 
\begin{equation}
\mathbf{H}=\mathbf{A}_{\tau}\mathbf{X}\mathbf{A}_{\theta}^{H},\label{eq:channel_delay_angle_decomposition}
\end{equation}
where
\begin{equation}
\mathbf{X}\triangleq\sum_{p=0}^{L-1}\rho_{p}\mathbf{e}_{k_{p},D}\mathbf{e}_{l_{p},M}^{T}\in\mathbb{C}^{D\times M},\label{eq:Wongrid}
\end{equation}
with $\mathbf{e}_{n,N}\in\mathbb{C}^{N\times1}$ denoting the canonical
basis vector with the $n$-th element equal to $1$. Matrix $\mathbf{X}$
is the \emph{delay-angular representation} of the channel, which is
a sparse matrix with $L$ nonzero elements out of a total $DM$. An
example of $\mathbf{X}$ with on-grid channel parameters is shown
in Figure \ref{fig grid and off-grid W} (left panel). This sparsity
of $\mathbf{X}$ (or its corresponding covariance matrix) has been exploited in the literature for obtaining
efficient channel estimators \cite{Swindlehurst2016_TSP,Heath JSAC 2017,Haghighatshoar2017_TWC}
by direct application of algorithms from the field of CS. However,
as will be argued in the following, the sparsity pattern (support)
of $\mathbf{X}$ is not completely random but follows a 
hierarchical pattern, a property that will be exploited for algorithm
design and rigorous analysis in terms of performance and overhead
required to achieve it. 

\section{Multiuser Channel Estimation Problem Statement}

Towards reducing the pilot overhead, the BS partitions the uplink
UEs to groups of $U$ UEs. Each group is assigned an exclusive set
of pilot subcarriers and all $V\leq U$ \emph{active} UEs within a
group transmit their pilots on these subcarriers and on the same OFDM
symbol. {For the analysis and design purposes, we consider an arbitrary UE group and discuss later the joint assignment of subcarriers to multiple groups. Let $\mathcal{N}_{p}\subseteq[N]$
denote the set of $N_{p}\triangleq|\mathcal{N}_{p}|$ dedicated
pilot subcarriers to this group.} Towards reducing implementation complexity, only
the received signals from a set $\mathcal{M}_{p}\subseteq[M]$ of
$M_{p}\triangleq|\mathcal{M}_{p}|$ antennas are considered at the BS for channel
estimation purposes. 

Let $\mathbf{P}_{\mathcal{N}_{p}}\triangleq\mathbf{I}_{N,\mathcal{N}_{p}}\in\{0,1\}^{N_{p}\times N}$
and $\mathbf{P}_{\mathcal{M}_{p}}\triangleq\mathbf{I}_{M,\mathcal{M}_{p}}\in\{0,1\}^{M_{p}\times M}$
denote the \emph{sampling} matrices in frequency and space, respectively.
The task of the BS is to identify all the UE channels from the observation
\begin{equation}
\mathbf{Y}=\sum_{u=0}^{U-1}\text{diag}(\mathbf{c}_{u})\mathbf{P}_{\mathcal{N}_{p}}\mathbf{H}_{u}\mathbf{P}_{\mathcal{M}_{p}}^{T}+\mathbf{Z}\in\mathbb{C}^{N_{p}\times M_{p}},\label{eq:observed_signal}
\end{equation}
where $\mathbf{c}_{u}\in\mathbb{C}^{N_{p}},\mathbf{H}_{u}\in\mathbb{C}^{N\times M}$,
are the \textcolor{black}{pilot} \emph{signature} and channel transfer
matrix of the $u$-th UE, respectively, and $\mathbf{Z}\in\mathbb{C}^{N_{p}\times M_{p}}$
is a noise matrix of arbitrary distribution apart from the mild assumption
that $\|\mathbf{Z}\|$ is finite with probability $1$. The elements
of $\mathbf{c}_{u}$ are known to the BS and assumed, w.l.o.g., to
be of unit modulus for all $u\in[U]$. For the $U-V$ UEs that are
not active, the channel transfer matrix is equal to an all-zeros matrix.
The receiver is not aware which UEs are inactive but does know $V$
as well as the number of channel paths $L$, assumed to be the same
for all UEs.

It follows from the discussion of Sec. II that the problem of estimating
the transfer matrices $\{\mathbf{H}_{u}\}_{u\in[U]}$ can be equivalently
posed as the problem of estimating the delay-angular channel representations
$\{\mathbf{X}_{u}\}_{u\in[U]}$. Setting $\mathbf{H}_{u}=\mathbf{A}_{\tau}\mathbf{X}_{u}\mathbf{A}_{\theta}^{H}$
in (\ref{eq:observed_signal}) and normalizing for technical reasons
by $1/\sqrt{N_{p}M_{p}}$ results in the system equation
\begin{equation}
\mathbf{Y}=\bar{\mathbf{A}}_{\tau}\bar{\mathbf{X}}\bar{\mathbf{A}}_{\theta}^{H}+\mathbf{Z},\label{eq:system model w.r.t. =00005Cbar(X)}
\end{equation}

\noindent where
\begin{equation}
\bar{\mathbf{A}}_{\tau}\triangleq\frac{1}{\sqrt{N_{p}}}\left[\text{diag}(\mathbf{c}_{0})\mathbf{P}_{\mathcal{N}_{p}}\mathbf{A}_{\tau},\ldots,\text{diag}(\mathbf{c}_{U-1})\mathbf{P}_{\mathcal{N}_{p}}\mathbf{A}_{\tau}\right],\label{eq:A_tau structure}
\end{equation}
\begin{equation}
\mathbf{\bar{\mathbf{A}}_{\theta}\triangleq}\frac{1}{\sqrt{M_{p}}}\mathbf{P}_{\mathcal{M}_{p}}\mathbf{A}_{\theta},\label{eq:A_theta structure}
\end{equation}
\[
\bar{\mathbf{X}}\triangleq\left[\mathbf{X}_{0}^{T},\mathbf{X}_{1}^{T},\ldots,\mathbf{X}_{U-1}^{T}\right]^{T},
\]

\noindent and, with a slight abuse of notation, we denote also by $\mathbf{Y}$ and 
$\mathbf{Z}$ the normalized observation and noise matrix, respectively. Note that $\bar{\mathbf{X}}$
is a sparse matrix with $VL$ non-zero elements out of a total $UDM$.

Towards expressing the linear model of (\ref{eq:system model w.r.t. =00005Cbar(X)})
in standard form (w.r.t. the unknown elements of $\bar{\mathbf{X}}$),
the matrix observation should be vectorized. Note that there are two
options to do this: Either consider $\text{vec}(\mathbf{Y})$ or $\text{vec}(\mathbf{Y}^{T})$,
which will be referred to as the \emph{frequency-space} (F-S) and
\emph{space-frequency} (S-F) option, respectively. These two options
are, of course, mathematically equivalent, when $\bar{\mathbf{X}}$
is treated as an arbitrary matrix. However, as the support of $\bar{\mathbf{X}}$
reflects physical channel properties, these two options suggest different
(additional) channel modeling assumptions, which can be algorithmically
exploited and result in different overhead requirements, as will be
discussed in the next section. 

By straightforward algebra, the channel estimation problem can be
stated as follows.
\begin{problem}
Find a computationally efficient estimator of $\mathbf{x}\in\mathbb{C}^{UDM}$
given the measurement 
\begin{equation}
\mathbf{y}=\mathbf{A}\mathbf{x}+\mathbf{z}\in\mathbb{C}^{N_{p}M_{p}},\label{eq:linear model}
\end{equation}

\noindent where $\mathbf{z}\in\mathbb{C}^{N_{p}M_{p}}$ is a noise
vector, and, under the F-S option,
\begin{equation}
\left\{ \begin{array}{c}
\mathbf{y}\triangleq\text{vec}(\mathbf{Y})\\
\mathbf{A}\triangleq\bar{\mathbf{A}}_{\theta}^{*}\otimes\bar{\mathbf{A}}_{\tau}\\
\mathbf{x}\triangleq\text{vec}(\bar{\mathbf{X}})
\end{array}\right\} ,\label{eq:option 1 system model}
\end{equation}

\noindent or, under the S-F option,
\begin{equation}
\left\{ \begin{array}{c}
\mathbf{y}\triangleq\text{vec}(\mathbf{Y}^{T})\\
\mathbf{A}\triangleq\bar{\mathbf{A}}_{\tau}\otimes\bar{\mathbf{A}}_{\theta}^{*}\\
\mathbf{x}\triangleq\text{vec}(\bar{\mathbf{X}}^{T})
\end{array}\right\} .\label{eq: option 2 system model}
\end{equation}
We also ask for the design (selection) of $\mathcal{N}_{p}$, $\mathcal{M}_{p}$,
and $\{\mathbf{c}_{u}\}_{u\in[U]}$, towards minimizing the pilot
subcarriers $N_{p}$ and observed antenna signals $M_{p}$ required
for reliable (in a specific sense that we will made
precise later) estimation.
\end{problem}
With $N_{p}M_{p}<UDM$, which is the case of interest in a wideband,
massive MIMO setting, the linear estimation problem of (\ref{eq:linear model})
becomes under-determined. However, as $\mathbf{x}$ is $VL$-sparse,
one can utilize tools from CS theory \cite{MathIntroToCS} for its
estimation. In particular, it is known that, in the absence of noise,
a necessary requirement for perfect recovery of $\mathbf{x}$ from
$\mathbf{y}$ (by means of any algorithm) is \cite[Theorem 11.6]{MathIntroToCS}
\begin{equation}
N_{p}M_{p}=\mathcal{O}\left(VL\log(UDM)\right)\text{ for }UDM\rightarrow\infty.\label{eq:universal overhead bound of CS recovery}
\end{equation}

\noindent We will refer to the product $N_{p}M_{p}$ as \emph{overhead}.
Equation (\ref{eq:universal overhead bound of CS recovery}) reveals
that the necessary overhead scales much slower than the overhead
corresponding to a naive consideration of all $N$ subcarriers and
$M$ antennas for channel estimation.

However, achievability of the universal bound of (\ref{eq:universal overhead bound of CS recovery})
depends crucially on the \emph{sensing matrix} $\mathbf{A}$ that
appears in (\ref{eq:linear model}). In particular, a typical sufficient condition for $\mathbf{A}$ is to satisfy the restricted isometry property (RIP)
(see Definition \ref{def: standard RIP definition}). This would indeed
be the case (with high probability) if the elements of $\mathbf{A}$
were, e.g., Gaussian distributed\textcolor{black}{{} }\cite{MathIntroToCS}\textcolor{black}{,
allowing the use of standard algorithms from CS theory for the recovery
of }$\mathbf{x}$ with the overhead of (\ref{eq:universal overhead bound of CS recovery}).\textcolor{black}{{}
Unfortunately, there is very limited flexibility in designing the
sensing matrix }$\mathbf{A}$ as the latter has by default the Kronecker
product structure shown in (\ref{eq:option 1 system model}) and (\ref{eq: option 2 system model})
and the design of the UE signatures only affects the constituent matrix
$\bar{\mathbf{A}}_{\tau}$ under the specific block structure of (\ref{eq:A_tau structure}).

{Works towards a characterization of the RIP constant of Kronecker-product sensing matrices are available \cite{Jokar Kronecker, Duarte Kronecker}, with the main result being a lower bound of in terms of the RIP constants of the individual constituent matrices. This bound can be used to obtain insights on the necessary (but not sufficient) scaling of training overhead. However, as will be shown later (cf. discussion after Theorem 9), this scaling is overly pessimistic. This is due to the fact that  $\mathbf{x}$ is not simply sparse, as treated by the standard CS approach, but \emph{hierarchically sparse}, a notion we define next, which effectively implies a reduction of the solution space for the channel estimation problem. This solution space reduction implies that the estimation problem is ``easier'' than the one implied by the standard CS treatment, hence a smaller training overhead is expected. In the following section, a family of recovery algorithms (in the presence of noise) exploiting the hierarchical sparsity are presented, for which rigorous scaling laws for the required training overhead are obtained based on the concept of \emph{Hierarchical RIP} (HiRIP).}

\section{Algorithm Design and Analysis Exploiting \textcolor{black}{Hierarchical
Sparsity}}

This section identifies important structural properties of $\mathbf{x}$
(under both F-S and S-F options), which are taken into account for
the design of efficient channel estimation algorithms as well as providing
performance guarantees. The latter will in turn provide design criteria
for the pilot signatures. For a specific pilot signature design, a
rigorous identification of the overhead scaling sufficient to guarantee
channel identification with bounded error is provided, which also
suggests that the F-S option is preferable towards minimum number
of pilots $N_{p}$.

\subsection{Hierarchical Sparsity Under F-S and S-F Options}

\begin{figure}[t]
\centering \includegraphics[scale=0.8]{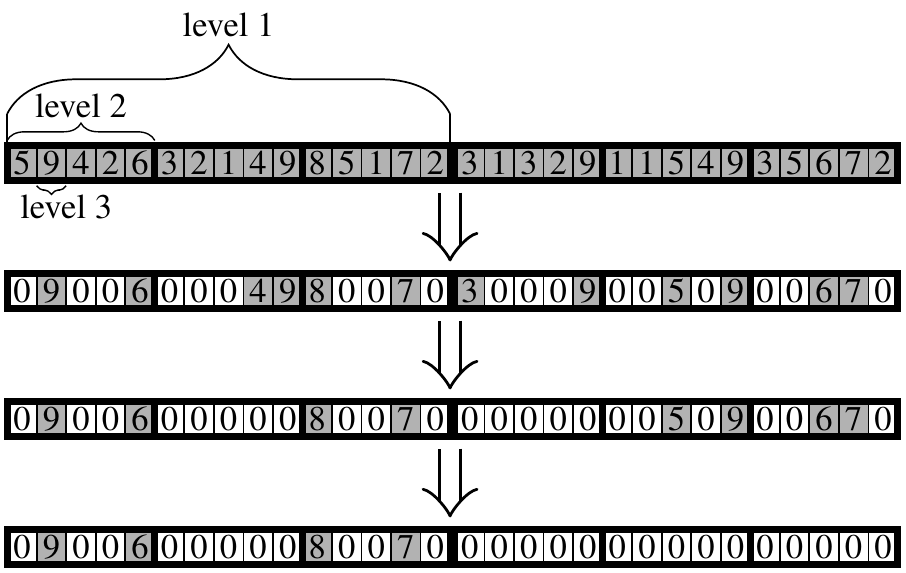}\caption{\label{fig:T_operator}{Illustration of the sequence of actions of
the $\mathcal{T}_{(1,2,2)}(\cdot)$ operator (as described in Algorithm
\ref{alg:T_operator_alg}) on a three-level block vector in $\mathbb{C}^{2\cdot 3 \cdot 5}$ ($2$ blocks
of $3$ blocks of $5$ elements each). The support of the best $(1,2,2)$-Hi-sparse
approximation of the vector is $\{1,4,10,13\}$, with index $0$ corresponding to the leftmost element in the vector. Note that this support is
different from the support $\{1,9,19,24\}$ of the best $1\times2\times2=4$-sparse approximation, when the vector is treated as an arbitrary vector (no block structure) in $\mathbb{C}^{30}$. }}
\end{figure}

The fundamental observation towards an efficient channel estimation
algorithm and rigorous performance analysis is that $\mathbf{x}$
is not only sparse, but its support possesses a certain structure,
called \emph{hierarchical sparsity } {\cite{C-HiLasso, Bockelmann,  Roth2016_TSP,RothEtAl2018_ISIT}}.
\begin{defn}[Hierarchical sparsity]
Let $\mathbf{s}=(s_{1},\dots,s_{\ell})$ be an $\ell$-tuple of natural
numbers and consider an $\ell$-level block vector $\tilde{\mathbf{x}}\in\mathbb{C}^{N_{1}\cdot N_{2}\cdots N_{\ell}}$,
with $N_{i}\geq s_{i},i\in\{1,2,\ldots,\ell\}$. We say that $\tilde{\mathbf{x}}$
is \emph{$\mathbf{s}$-hierarchically-sparse} (written as $\mathbf{s}$-Hi-sparse)
if it has the property of hierarchical $\mathbf{s}$-sparsity defined
inductively as follows: For $\ell=1$, $\tilde{\mathbf{x}}$ is $\mathbf{s}$-Hi-sparse
if at most $s_{1}$ of its $N_{1}$ elements are non-zero (this is
the standard notion of sparsity). For $l>1$, $\tilde{\mathbf{x}}$
is called $\mathbf{s}$-Hi-sparse if it consists of $N_{1}$ blocks
and at most $s_{1}$ of these are non-zero with each non-zero block
being $(s_{2},\dots,s_{\ell})$-Hi-sparse.\textcolor{red}{{} }The lower
part of Fig. \ref{fig:T_operator} demonstrates an example of a vector
in $\mathbb{C}^{2\cdot3\cdot5}$ that is $(1,2,2)$-Hi-sparse.
\end{defn}

 {It is noted that the notion of hierarchical sparsity is more general than that of the common block sparsity where a vector of length $N$ is partitioned into $N/d$ blocks of $d$ elements each, with $s<N/d$ of the blocks (and their elements) being non-zero \cite{block sparse}. Note that in this case, the vector can be treated as a two-level block vector in $\mathbb{C}^{\frac{N}{d} \cdot d}$ that is $(s,d)$-Hi-sparse.}

It is easy to see that the unknown vector $\mathbf{x}$ in (\ref{eq:linear model})
is actually a hierarchically sparse, $3$-level block vector under
both F-S and S-F options. In particular, under the F-S option and
the assumption $LV\leq M$ (reasonable for massive MIMO and sparse
channels), $\mathbf{x}\in\mathbb{C}^{M\cdot U\cdot D}$ and is $(LV,V,L)$-Hi-sparse.
Note that the first (outer) hierarchy level corresponds to angles
(up to $LV$ angle values can be present, equal to the number of total
paths from all active UEs), the second hierarchy level corresponds
to UEs (up to $V$ active UEs can have a path with the same angle),
and the third hierarchy level corresponds to delays (up to $L$ delays
per UE per angle can be present, equal to the total paths per UE).
However, for (asymptotically) large $M$, one may reasonably assume
that (a) for each angle value there can be no more than $K_{V}<V$
UEs with a channel path having this angle and (b) for each angle there
can be no more than $K_{L}<L$ paths for each UE with this value,
rendering $\mathbf{x}$ as $(LV,K_{V},K_{L})$-Hi-sparse. Under the
S-F option, $\mathbf{x}\in\mathbb{C}^{U\cdot D\cdot M}$ and is $(V,L,L)$-Hi-sparse,
with the first (outer) hierarchy level corresponding to UEs ($V$
out of $U$ UEs active), the second corresponding to delays (up to
$L$ paths present per UE), and the third corresponding to angles
(up to $L$ paths with the same delay). Similar to the F-S option,
the hierarchical sparsity characterization under S-F option can be
refined in the (asymptotically) large $N$ regime, where up to $K_{L}$
paths can be assumed to have the same delay per UE, rendering $\mathbf{x}$
as $(V,L,K_{L})$-Hi-sparse. 

The F-S and S-F options result in a different ordering of
levels for $\mathbf{x}$ and suggest different (but reasonable) assumptions
for the delay/angular distribution of UE channels. However, at this
point, it is not clear which of the two options is preferable. Note also that, 
for asymptotically large $M$ or $N$, it is reasonable to assume
that \emph{$K_{V}=K_{L}=1$}, although we do not explicitly write them as such for generality of presentation. 

\subsection{Algorithm Design}

Clearly, the hierarchically sparse property of $\mathbf{x}$ \emph{should
be exploited} in algorithm design and analysis as it provides significant
restrictions on its support, compared to the standard notion of sparsity
(which would characterize $\mathbf{x}$ simply as $VL$-sparse). \textcolor{black}{Towards
this end, the low-complexity, iterative hard thresholding (IHT) and
hard threshold pursuit (HTP) algorithms \cite{MathIntroToCS} are
modified as shown in Algorithm \ref{alg:HiIHT/HiHTP-Channel-Estimation}
to take into account the hierarchical sparsity }of $\mathbf{x}$\textcolor{black}{{}
and are referred to in the following as hierarchical IHT (HiIHT) and
hierarchical HTP (HiHTP), respectively. The algorithms can be applied
equally well under either the F-S or S-F option and are independent
of the noise statistics.}

\begin{algorithm}[tbh]
\caption{HiIHT/HiHTP Channel Estimation\label{alg:HiIHT/HiHTP-Channel-Estimation}}
\begin{algorithmic}[1]
\REQUIRE $\mathbf{y}$, $\mathbf{A}$, $V$, $L$, $K_L$, $K_V$ (the latter only under F-S option).
\STATE $i=0$, $\hat{\mathbf{x}}^{\left(  0\right)  }=\mathbf{0} \in \mathbb{C}^{UDM}$
\REPEAT
\STATE $i = i + 1$, 
\STATE$\hat{\mathbf{x}}_\text{temp} =  \hat{\mathbf{x}}^{\left(  i-1\right)  }+\mathbf{A}^{H}\left(  {\mathbf{y}}-\mathbf{A} \hat{\mathbf{x}}^{\left(  i-1\right)  }\right)   $
\STATE$\hat{\mathcal{S}}^{\left(  i\right)  }=
\begin{cases}        
\mathcal{T}_{(VL,K_V,K_L)}(\hat{\mathbf{x}}_\text{temp}), &\text{F-S option,}\\         
\mathcal{T}_{(V,L,K_L)}(\hat{\mathbf{x}}_\text{temp}), &\text{S-F option}
\end{cases}$
\IF{HiIHT}
\STATE $\hat{\mathbf{x}}^{\left(  i\right)  }=\mathbf{0} \in \mathbb{C}^{UDM}$
\STATE$\hat{\mathbf{x}}^{\left(  i\right)  }_{\hat{\mathcal{S}}^{\left(  i\right)  }}=\hat{\mathbf{x}}_{\text{temp},\hat{\mathcal{S}}^{\left(  i\right)  }}  $ \ELSIF{HiHTP} 
\STATE$\hat{\mathbf{x}}^{\left(  i\right)  }=\arg\min_{\beta\in 	\mathbb{C}^{UDM}, \text{supp}({\beta})\subseteq \hat{\mathcal{S}}^{\left(  i\right) }} \left\{  \Vert\mathbf{y}-\mathbf{A} \beta \Vert\right\}  $ 
\ENDIF
\UNTIL stopping criterion is met at $i=i^{\ast}$
\RETURN$\begin{cases}        
(VL, K_V, K_L)\text{-Hi-sparse }\hat{\mathbf{x}}^{\left(  i^{\ast}\right)  }, &\text{F-S option,}\\         
(V, L, K_L)\text{-Hi-sparse }\hat{\mathbf{x}}^{\left(  i^{\ast}\right)  }, &\text{S-F option}
\end{cases}$
\end{algorithmic} 

\end{algorithm}
In iteration $i$, the estimate of iteration $i-1$ is first updated
by a standard gradient-descent step to obtain $\hat{\mathbf{x}}_{\text{temp}}$.
From $\hat{\mathbf{x}}_{\text{temp}}$, the hierarchically sparse
support $\mathcal{S}\in[UDM]$ of $\mathbf{x}$ is estimated by application
of the \emph{thresholding} operator $\mathcal{T}_{(\cdot,\cdot,\cdot)}(\cdot)$, to be defined next.
For HiIHT, the current iteration estimate of $\mathbf{x}$ is set
equal to $\hat{\mathbf{x}}_{\text{temp}}$ except for its elements
that do not belong to the estimated support and are set equal to zero.
 {For HiHTP, the iteration $i$ estimate is obtained as the hierarchically sparse vector whose non-zero element values are obtained by  minimizing a standard least squares cost function.}

Utilization of the operator $\mathcal{T}_{(\cdot,\cdot,\cdot)}(\cdot)$
is the only but critical differentiator of HiIHT/HiHTP compared to
their ``standard'' IHT/HTP counterparts \cite{MathIntroToCS}. In
particular, for any multi-level block vector $\tilde{\mathbf{x}}\in\mathbb{C}^{N_{1}\cdot N_{2}\cdots N_{\ell}}$
and any $\mathbf{s}=(s_{1},s_{2},\ldots,s_{\ell})$, $\mathcal{T}_{\mathbf{s}}(\tilde{\mathbf{x}})$
is defined as the support of the multi-level block vector $\tilde{\mathbf{z}}\in\mathbb{C}^{N_{1}\cdot N_{2}\cdots N_{\ell}}$
that is $\mathbf{s}$-Hi-sparse and minimizes $\|\tilde{\mathbf{x}}-\tilde{\mathbf{z}}\|$.
Its action can be computed recursively with minimal complexity
as described in Algorithm \ref{alg:T_operator_alg} with an example
of this computation shown in Fig. \ref{fig:T_operator}.  {For the  channel estimation problem, $\ell=3$ in the description of the algorithm, with $(N_1,N_2,N_3)=(M,U,D)$ and $(N_1,N_2,N_3)=(U,D,M)$ under the F-S and S-F option respectively.}

\begin{algorithm}[tbh]
\caption{Action of operator $\mathcal{T}_{\mathbf{s}}(\cdot)$\label{alg:T_operator_alg}}
\begin{algorithmic} 		[1] 	
\REQUIRE $\tilde{\mathbf{x}} \in \mathbb{C}^{N_1 \cdot N_2 \cdots N_\ell}$, $\mathbf{s}=(s_1,s_2,\ldots, s_\ell)$, $\ell \geq 2$ 
\STATE $\tilde{\mathbf{z}} = \tilde{\mathbf{x}}$.	
\STATE For each of the $N_1 N_2 \cdots N_{\ell-1}$ blocks at level $\ell-1$ of $\tilde{\mathbf{z}}$, identify the $s_\ell$ (out of a total $N_\ell$) largest-modulus elements  and set the remaining elements equal to zero. Ties are resolved arbitrarily.	
\STATE 		$k=\ell-2$ .		
\WHILE{$k\geq 1$} 		
\STATE For each of the $N_1 N_2 \cdots N_k$ blocks at level $k$ of $\tilde{\mathbf{z}}$, identify the $s_{k+1}$ (out of a total $N_{k+1}$) blocks with the largest Euclidean norm and set the elements of the remaining blocks equal to zero. Ties are resolved arbitrarily.		
\STATE 		$k=k-1$ 		
\ENDWHILE 		
\RETURN $\text{supp}(\tilde{\mathbf{z}})$
\end{algorithmic} 
\end{algorithm}

\subsection{Performance Analysis and Overhead Requirements}

Towards characterizing the performance of HiIHT/HiHTP, which, in turn,
will provide insights on pilot signature design and overhead requirements,
the concept of \emph{hierarchical} \emph{RIP} (HiRIP) constant, first
introduced in \cite{Roth2016_TSP}, is essential. 
\begin{defn}[HiRIP constant]
 Let $\mathbf{s}=(s_{1},s_{2},\ldots,s_{\ell})$ be an $\ell$-tuple
of natural numbers. The $\mathbf{s}$-HiRIP constant $\delta_{\mathbf{s}}(\tilde{\mathbf{A}})$
of a (deterministic) matrix $\tilde{\mathbf{A}}\in\mathbb{C}^{N_{0}\times(N_{1}N_{2}\cdots N_{\ell})}$
is the smallest $\delta\geq0$ such that 
\begin{equation}
(1-\delta)\Vert\tilde{\mathbf{x}}\Vert^{2}\leq\Vert\tilde{\mathbf{A}}\tilde{\mathbf{x}}\Vert^{2}\leq(1+\delta)\Vert\tilde{\mathbf{x}}\Vert^{2},\label{eq:RIPequation}
\end{equation}
for all $\mathbf{s}$-Hi-sparse $\ell$-level block vectors $\tilde{\mathbf{x}}\in\mathbb{C}^{N_{1}\cdot N_{2}\cdots N_{\ell}}$.
We say that $\tilde{\mathbf{A}}$ satisfies the $\mathbf{s}$-HiRIP
if $\delta_{\mathbf{s}}(\tilde{\mathbf{A}})<\bar{\delta}$ where $\bar{\delta}<1$ is a pre-specified constant.\footnote{Note the difference in the notation $\delta_s(\cdot)$ and $\delta_\mathbf{s}(\cdot)$ for the RIP and HiRIP constants, respectively. A scalar $s$ is used as a subscript for RIP, whereas a vector $\mathbf{s}$, sometimes with its elements explicitly indicated, is used for HiRIP.}
\end{defn}
\begin{rem}
\label{rem:HiRIP stricter than RIP}The definition of the $\mathbf{s}$-HiRIP
constant closely follows Def. \ref{def: standard RIP definition}
of the (standard) $s$-RIP constant and they actually coincide when
$\mathbf{s}$ contains only a single element, i.e., it is a scalar.
However, when $\mathbf{s}$ is a vector of two or more elements, the
notion of HiRIP constant is not directly comparable to that of the
RIP constant as the first applies to hierarchically sparse vectors
whereas the second applies to more general, sparse vectors (that may
or may not be hierarchically sparse). However, a link between the
two notions exists by noting that, for any matrix $\tilde{\mathbf{A}}$,
it must hold  {(see Appendix \ref{sec:proof of Eq 14})}
\begin{equation}
\delta_{(s_{1},s_{2},\ldots,s_{\ell})}(\tilde{\mathbf{A}})\leq\delta_{s_{1}s_{2}\cdots s_{\ell}}(\tilde{\mathbf{A}}),\label{eq:HIRIP<=00003DRIP}
\end{equation}
for any $s_{1},s_{2},\ldots,s_{\ell}$, a result that will be utilized
in the following.\textcolor{red}{{} }
\end{rem}
The HiRIP framework allows to obtain the following rigorous guarantees
for the performance of HiIHT/HiHTP.
\begin{thm}[Recovery guarantee of HiIHT/HiHTP]
\label{thm:HiHTP_performance}Assume that $M,N,D\gg L$ and suppose
that the sensing matrix $\mathbf{A}$ in (\ref{eq:linear model})
has a HiRIP constant
\[
\delta\triangleq\begin{cases}
\delta_{(3LV,3K_{V},3K_{L})}(\mathbf{A}), & \mathcal{\text{\emph{under F-S option}}},\\
\delta_{(3V,3L,3K_{L})}(\mathbf{A}), & \mathcal{\text{\emph{under S-F option,}}}
\end{cases}
\]

\noindent with
\begin{equation}
\delta<1/\sqrt{3}.\label{eq:HiRIP requirement}
\end{equation}

\noindent Then, the sequence of estimates $\{\hat{\mathbf{x}}^{\left(i\right)}\}$
generated by the HiIHT and HiHTP algorithms satisfies
\[
\Vert\mathbf{x}-\hat{\mathbf{x}}^{\left(i\right)}\Vert\leq\kappa^{i}\Vert\mathbf{x}\Vert+\tau\Vert\mathbf{z}\Vert,
\]

\noindent for all $i\geq0$, with 
\[
\kappa\triangleq\begin{cases}
\sqrt{3}\delta, & \text{\emph{for HiIHT},}\\
\sqrt{2\delta/(1-\delta^{2})}, & \text{\emph{for HiHTP},}
\end{cases}
\]

\noindent and 
\[
\tau\triangleq\begin{cases}
2.18/(1-\kappa), & \text{\emph{for HiIHT}},\\
5.15/(1-\kappa), & \emph{for HiHTP}.
\end{cases}
\]
\end{thm}
\begin{IEEEproof}
The result for the HiHTP follows directly from application of \cite[Theorem 4]{Roth2016_TSP}.
The proof for the HiIHT follows the HiHTP proof with the same modifications
as the ones considered in the recovery guarantee proofs of the HTP/IHT
algorithms given in \cite[Theorem 6.18]{MathIntroToCS}.
\end{IEEEproof}
It follows that in order to ensure \emph{reliable} channel estimation
in the sense of perfect and bounded-error recovery of $\mathbf{x}$
via the HiHTP/HiIHT algorithms in the noiseless ($\|\mathbf{z}\|=0$)
and noisy ($\|\mathbf{z}\|>0$) case, respectively, we need to design
$\mathcal{N}_{p}$, $\mathcal{M}_{p}$, and $\{\mathbf{c}_{u}\}_{u=0}^{U-1}$
such that the HiRIP constant of $\mathbf{A}$ satisfies (\ref{eq:HiRIP requirement}).
Similar to RIP, the explicit computation of HiRIP constants is a very
difficult problem (even numerically) \cite{RothEtAl2018_ISIT}.
However, the following bound on the HiRIP constant of a Kronecker-product
sensing matrix in terms of the RIP constants of its factor matrices
is available, which can be used to obtain a rigorous description of
the overhead required to achieve (\ref{eq:HiRIP requirement}).
\begin{lem}
\label{thm:HiRIP bound}Consider a matrix $\tilde{\mathbf{A}}\triangleq\tilde{\mathbf{A}}_{1}\otimes\tilde{\mathbf{A}}_{2}$,
with $\tilde{\mathbf{A}}_{k}\in\mathbb{C}^{M_{k}\times N_{k}},k=1,2$,
which, for all $3$-level block vectors $\tilde{\mathbf{x}}\in\mathbb{C}^{N'_{1}\cdot N_{2}'\cdot N_{3}'}$
with $N_{1}'N_{2}'N_{3}'=N_{1}N_{2}$, has an $\mathbf{s}$-HiRIP
constant $\delta_{\mathbf{s}}(\tilde{\mathbf{A}})$ for some $\mathbf{s}\triangleq(s_{1},s_{2},s_{3})$.
If $N_{1}=N_{1}'$ and $N_{2}=N_{2}'N_{3}'$, it holds
\begin{equation}
\delta_{\mathbf{s}}(\tilde{\mathbf{A}})\leq\left(1+\delta_{s_{1}}(\tilde{\mathbf{A}}_{1})\right)\left(1+\delta_{s_{2}s_{3}}(\tilde{\mathbf{A}}_{2})\right)-1,\label{eq:HiRIP bound case 1}
\end{equation}

\noindent whereas, if $N_{1}=N_{1}'N_{2}'$ and $N_{2}=N_{3}'$, it
holds
\begin{equation}
\delta_{\mathbf{s}}(\tilde{\mathbf{A}})\leq\left(1+\delta_{s_{1}s_{2}}(\tilde{\mathbf{A}}_{1})\right)\left(1+\delta_{s_{3}}(\tilde{\mathbf{A}}_{2})\right)-1.\label{eq:HiRIP bound case 2}
\end{equation}
\end{lem}
\begin{IEEEproof}
The bound of (\ref{eq:HiRIP bound case 1}) follows from the inequality
\cite[Theorem 4]{RothEtAl2018_ISIT}
\[
\delta_{\mathbf{s}}(\tilde{\mathbf{A}})\leq\left(1+\delta_{s_{1}}(\tilde{\mathbf{A}}_{1})\right)\left(1+\delta_{(s_{2},s_{3})}(\tilde{\mathbf{A}}_{2})\right)-1,
\]

\noindent and (\ref{eq:HIRIP<=00003DRIP}). The bound of (\ref{eq:HiRIP bound case 2})
follows by the inequality 
\[
\delta_{\mathbf{s}}(\tilde{\mathbf{A}})\leq\left(1+\delta_{(s_{1},s_{2})}(\tilde{\mathbf{A}}_{1})\right)\left(1+\delta_{s_{3}}(\tilde{\mathbf{A}}_{2})\right)-1,
\]

\noindent which can be shown to hold by a straightforward extension
of the proof of \cite[Theorem 4]{RothEtAl2018_ISIT} and again applying
(\ref{eq:HIRIP<=00003DRIP}).
\end{IEEEproof}

The importance of this theorem is that it bounds the HiRIP constant
of $\mathbf{A}$ in terms of the RIP constants of its constituent
matrices $\bar{\mathbf{A}}_{\tau}$ and $\bar{\mathbf{A}}_{\theta}^{*}$.
 {Any design resulting in this bound of the HiRIP constant of $\mathbf{A}$ been less than $1/\sqrt{3}$} is therefore sufficient to achieve the performance guarantees of Theorem
\ref{thm:HiHTP_performance}. To this end, we propose the following
design. 
\begin{defn}[System Design]
Set $U\leq N/D$ and let $\mathbf{c}\in\mathbb{C}^{N}$ be an arbitrary
sequence of unit modulus elements. For an arbitrary group of $U$
UEs, the set of its dedicated pilot subcarriers $\mathcal{N}_{p}$
is a randomly and uniformly selected subset of $[N]$ with cardinality
$N_{p}$, whereas the set of observed antennas $\mathcal{M}_{p}$
is randomly and uniformly selected subset of $[M]$ with cardinality
$M_{p}$ (same for all UE groups). The UE signature sequences are
\begin{equation}
\mathbf{c}_{u}=\mathbf{P}_{\mathcal{N}_{p}}\text{diag}\left(\left[1,e^{-j\frac{2\pi}{N}uD},\ldots,e^{-j\frac{2\pi}{N}uD(N-1)}\right]\right)\mathbf{c},\label{eq: frequency_shifted_signatures}
\end{equation}
for all $u\in[U].$ 

 {Note that under this design the joint assignment of pilot subcarriers to multiple UE groups is simplified to a random partition of subcarriers.} This design is motivated by the availability of rigorous RIP constant
characterization for matrices obtained by random sampling of rows
of orthogonal matrices. Indeed, it immediately follows from (\ref{eq:A_theta structure}) and 
the random selection of antennas that $\bar{\mathbf{A}}_{\theta}^{*}$
is a random sampling of the rows of the orthogonal matrix $(1/\sqrt{M_{p}})\mathbf{F}_{M,M}^{*}$,
whereas, direct substitution of (\ref{eq: frequency_shifted_signatures}) into (\ref{eq:A_tau structure}) results in
\begin{equation}
\mathbf{\bar{A}}_{\tau}=(1/\sqrt{N_{p}})\mathbf{P}_{\mathcal{N}_{p}}\text{diag}(\mathbf{c})\mathbf{F}_{N,UD},\label{eq:A_tau_with_proposed_design}
\end{equation}

\noindent i.e., $\mathbf{\bar{A}}_{\tau}$ is a random sampling of
the rows from the first $UD$ columns of the orthogonal matrix $(1/\sqrt{N_{p}})\text{diag}(\mathbf{c})\mathbf{F}_{N,N}$.
Equally important, this form of $\bar{\mathbf{A}}_{\theta}^{*}$ and
$\bar{\mathbf{A}}_{\tau}$ allows for the efficient computation of
the gradient-descent step in Algorithm \ref{alg:HiIHT/HiHTP-Channel-Estimation}
by means of fast Fourier transform (FFT).  {This makes HiIHT in particular especially attractive for application in systems with (very) large $M$ and/or $N$.} We note that phase-shifted
pilot sequence designs similar to (\ref{eq: frequency_shifted_signatures})
were also proposed in \cite{Leus optimal training MIMO OFDM,Swindlehurst2016_TSP,Al-Dhahir},
however, under different contexts in terms of system model and/or
assuming regularly-spaced pilot subcarriers.  {In addition, the well-known Zadoff-Chu sequences employed in cellular standards \cite{ZC_LTE} are compatible with the design of (\ref{eq: frequency_shifted_signatures}).}

The proposed design cannot be claimed to be optimal in the sense that it
is not obtained as the explicit solution of an optimization problem.
However, as will be shown in Sec. VI, it achieves very good performance
and, equally important, results in a sensing matrix $\mathbf{A}$ whose HiRIP can be analytically characterized as a function of   $N_p$ and $M_p$. This characterization, in combination with the performance guarantees of Theorem 6, allows for rigorous analytical insights on
the overhead requirements for reliable channel estimation, as stated in the following result.
\end{defn}
\begin{thm}
\label{thm:required overhead} Let $\delta_{\tau}>0$, $\delta_{\theta}>0$
be two arbitrary numbers that satisfy $\delta_{\tau}+\delta_{\theta}+\delta_{\tau}\delta_{\theta}<1/\sqrt{3}$.
With the proposed design and with a probability greater than $1-M^{-\log^{3}(M)}-N^{-\log^{3}(N)}$,
the HiIHT/HiHTP algorithm performance is as described in Theorem \ref{thm:HiHTP_performance}
when it holds
\begin{align}
N_{p} & \geq\min\left\{ 3C\delta_{\tau}^{-2}K_{V}K_{L}\log^{4}(N),N\right\} ,\label{eq:O_tau min}\\
M_{p} & \geq\min\left\{ 9C\delta_{\theta}^{-2}VL\log^{4}(M),M\right\} ,\label{eq:O_theta min}
\end{align}
under the F-S option, or 
\begin{align}
N_{p} & \geq\min\left\{ 9C\delta_{\tau}^{-2}VL\log^{4}(N),N\right\} ,\label{eq:O_tau min-1}\\
M_{p} & \geq\min\left\{ 3C\delta_{\theta}^{-2}K_{L}\log^{4}(M),M\right\} ,\label{eq:O_theta min-1}
\end{align}

\noindent under the S-F option, where $C>0$ is a universal constant.
\end{thm}
\begin{IEEEproof}
Please see Appendix \ref{sec:proof of required overhead}.
\end{IEEEproof}
The following remarks are in order:
\begin{itemize}
\item Both F-S and S-F options require an overhead $N_{p}M_{p}$ that is
proportional to $VL$ (assuming $M,N\gg L$ and $K_{V},K_{L}$ independent
of $V$ and $L$), similarly to the universal bound of (\ref{eq:universal overhead bound of CS recovery}).
Of course, using $N_{p}=N$ and $M_{p}=M$ will result in the best
performance in the presence of noise, however, this would be achieved
with an overly large overhead cost.
\item There is flexibility in distributing the overhead over the frequency
and space dimensions by changing the values of $\delta_{\tau}$ and
$\delta_{\theta}$ in Theorem \ref{thm:required overhead}. The minimum
pilot overhead ($N_p$) is achieved with $\delta_{\theta}=\epsilon$ and $\delta_{\tau}=1/\sqrt{3}-\epsilon$, for some arbitrarily small $\epsilon>0$,
resulting in $M_{p}=M$, i.e., all antennas are utilized, whereas
minimum number of observed antennas is achieved with $\delta_{\tau}=\epsilon$
and $\delta_{\theta}=1/\sqrt{3}-\epsilon$ with all subcarriers utilized for
training.
\item The scaling laws for $N_{p}$ and $M_{p}$ are different between the
F-S and S-F option due to the different hierarchical sparsity properties
for $\mathbf{x}$ corresponding to each of these (see discussion in
Sec. IV. A). Interestingly, under the F-S option and assuming that
$K_{V}$ and $K_{L}$ are independent of $V$ and $L$,\emph{ $N_{p}$
is independent of the number of channel paths $L$ and active UEs
$V$}, which is particularly appealing as it implies a robust pilot
design without a need for pilot reconfiguration with changing $L$
and/or $V$. Of course, one expects that performance will degrade
with increasing $L$ and/or $V$ with a fixed $N_{p}$, however, as
long as (\ref{eq:O_theta min}) holds, this degradation is expected
to be graceful in the sense of achieving a bounded estimation error,
as also verified in the numerical results of Sec. VI. Similar conclusions
hold for the S-F option, this time with $M_{p}$ being independent
of $L$ and $V$.
\item  {The result of  Theorem \ref{thm:required overhead}, even though only sufficient, provides a much better indication of the minimum possible overhead requirements than the one provided by conventional (unstructured) CS theory. Indeed, under the conventional CS treatment, a sufficient condition to achieve reliable channel estimation is $\delta_{cVL}(\mathbf{A})<\delta_{\text{RIP}}$, where the values of $c>0$ and $\delta_{\text{RIP}}>0$ depend on the considered estimation algorithm \cite{MathIntroToCS}. For Kronecker-type sensing matrices $\tilde{\mathbf{A}}=\tilde{\mathbf{A}}_1 \otimes \tilde{\mathbf{A}}_2$, it is known that $\delta_{s}(\tilde{\mathbf{A}}) \geq \max\{\delta_{s}(\mathbf{A}_1), \delta_{s}(\mathbf{A}_2) \}$, for any $s$ \cite{Jokar Kronecker, Duarte Kronecker}, which for the massive MIMO channel estimation problem implies that the pilot design should be such that it holds $\delta_{cVL}(\mathbf{A}_\tau)<\delta_{\text{RIP}}$ \emph{ and } $\delta_{cVL}(\mathbf{A}_\theta)<\delta_{\text{RIP}}$.
For the training sequence design considered above, it is easy to show that in order to achieve this condition \emph{both} $N_p$ and $M_p$ should scale proportionally to $VL$ (up to logarithmic factors), irrespective of $c$ and $\delta_{\text{RIP}}$. In contrast, Theorem 9 reveals that \emph{only one} of $N_p$ and $M_p$ needs to scale proportionally to $VL$ (up to logarithmic factors). }

\item The overhead requirement of Theorem \ref{thm:required overhead} is
 a sufficient condition  for the application of the HiIHT/HiHTP
algorithms. Therefore, this value may be greater than the necessary
and sufficient overhead requirement when a more sophisticated and
more complex algorithm such as, e.g., maximum likelihood estimation,
is employed.
\item The HiIHT/HiHTP algorithm description, performance analysis, and overhead
requirements described in this section are \emph{independent} of the
statistics of UE channels as well as noise. The only assumption considered
is that each UE channel consists of $L$ paths with on-grid values
for angles and delays.
\end{itemize}
As the pilot overhead reduction is critical towards increasing the
system capacity, i.e., accommodate more UEs and/or increase per-UE
rates, it is clear that the F-S option is preferable as $N_{p}$
does not scale with $V$ and $L$ (assuming that $K_V$ and $K_L$ are also independent of $V$, $L$). In particular, we have the following sufficient
pilot overhead requirement obtained by setting $\delta_{\theta}=\epsilon$
and $\delta_{\tau}=1/\sqrt{3}-\epsilon$, $\epsilon \rightarrow 0$, in Theorem \ref{thm:required overhead}.
\begin{cor}
Towards achieving reliable channel estimation with minimum pilot overhead,
the F-S option should be selected with $M_{p}=M$ (full antenna array
utilization) and
\[
N_{p}\geq CK_{V}K_{L}\log^{4}(N),
\]

\noindent where $C$ is a universal constant. 
\end{cor}
It is noted that the independence of the scaling behavior of the pilot
overhead from $L$ and $V$ is only possible by the utilization of
a massive number of antennas. In a loose sense, under the F-S option,
we shift the estimation burden to the spatial domain and corresponding
measurements, thus allowing for a minimum overhead in the frequency
domain. It is easy to see that when $M=1$, only the S-F option is
available, which results in a pilot overhead that scales with $L$
and $V$.

\section{Extension to Off-Grid Channel Parameters}

The previous sections considered on-grid channel parameters, which
can be assumed to be a good approximation in the regime of asymptotically
large $M$ and $N$. The fundamental benefit offered by this assumption
is that it naturally introduces the delay-angular channel representation
according to (\ref{eq:channel_delay_angle_decomposition}) and (\ref{eq:Wongrid})
that is exploited for algorithm development and system design. Considering
an arbitrary UE transfer matrix $\mathbf{H}\in\mathbb{C}^{N\times M}$
corresponding to a channel with off-grid parameters, a unique delay-angular
channel representation as in (\ref{eq:channel_delay_angle_decomposition})
exists only by treating the delay spread as equal to the OFDM symbol
duration $T_{s}$, i.e., $D=N$, even if the actual spread is actually
smaller than this value. Under this assumption, it follows from (\ref{eq:channelmatrix})
and (\ref{eq:channel_delay_angle_decomposition}) that the delay-angular
representation equals
\begin{align}
\mathbf{X} & =\mathbf{F}_{N,N}^{-1}\mathbf{H}(\mathbf{F}_{M,M}^H)^{-1}.\label{eq:delay-angular_representation_off_grid}
\end{align}

An example of $\mathbf{X}$ for a channel with off-grid parameters
is shown in Fig. \ref{fig grid and off-grid W} (right panel). It
can be seen that, in contrast to the on-grid case, the energy of each
path is leaked over \emph{all} elements of $\mathbf{X}$ rendering
it non-sparse. However, most of the energy of each path is concentrated
on a few elements of $\mathbf{X}$, suggesting that the latter can
be approximated by a sparse matrix, which, as in the on-grid case,
can be exploited in the channel estimation procedure. This approximate
sparsity of $\mathbf{X}$ in the off-grid case is confirmed by the
following result.
\begin{thm}
\label{thm:sparsification_in_off_grid_case}Let $L_{1}\leq\frac{N-1}{2}$,
$L_{2}\leq\frac{M-1}{2}$ be strictly positive integers. Setting $D=N$,
the delay-angle representation $\mathbf{X}\in\mathbb{C}^{N\times M}$
of any channel with $L$ paths of arbitrary (off-grid) delay and angle
values can be approximated by a sparse matrix $\mathbf{X}_{\text{\emph{sp}}}\in\mathbb{C}^{N\times M}$
that consists of at most $L(2L_{1}+1)(2L_{2}+1)$ non zero elements
with an error
\begin{equation}
\left\Vert \mathbf{X}-\mathbf{X}_{\text{\emph{sp}}}\right\Vert \leq\left(\frac{1}{\sqrt{L_{1}}}+\frac{1}{\sqrt{L_{2}}}\right)\sum_{p=0}^{L-1}\left|\rho_{p}\right|,\label{eq:off_grid_sparsification_error_bound}
\end{equation}

\noindent where $\rho_{p}$ is the complex gain of the $p$-th channel
path. 
\end{thm}
\begin{IEEEproof}
Please see Appendix \ref{sec:Proof_of_sparsification_if_off_grid_case}.
\end{IEEEproof}
The result implies that by choosing the parameters $L_{1}$ and $L_{2}$
sufficiently large, the delay-angular representation of any channel
with $L$ off-grid paths can be approximated with small error by the
delay-angular representation of a channel with $L(2L_{1}+1)(2L_{2}+1)$
on-grid paths. This increase of on-grid equivalent paths is due to
the, so called, basis mismatch error \cite{chi2011sensitivity} and
can be viewed as the cost of representing the channel on the fixed
basis corresponding to the dictionary matrices $\mathbf{A}_{\tau}$,
$\mathbf{A}_{\theta}$.

Since an accurate on-grid representation is available, the HiIHT/HiHTP
algorithms operating under the on-grid assumption can be employed
to identify the $L(2L_{1}+1)(2L_{2}+1)$ equivalent on-grid paths
per UE. In particular, the proof of Theorem \ref{thm:sparsification_in_off_grid_case}
considers a delay-angular representation where most of each path energy
spills over $L_{1}$ consecutive on-grid delay values and $L_{2}$
consecutive on-grid delay values. The example of Fig. \ref{fig grid and off-grid W}
identifies these energy regions for each path assuming $L_{1}=L_{2}=1$.
By the same arguments discussed in the on-grid case and considering
the F-S option, the sparse vector $\mathbf{x}$ to be estimated according
to the model (\ref{eq:linear model}) by HiIHT/HiHTP is now $(VL(2L_{2}+1),K_{V},K_{L}(2L_{1}+1))$-Hi-sparse
in $\mathbb{C}^{M\cdot U\cdot D}$.

Note that this approach will introduce the following four errors compared
to the on-grid case discussed in the previous sections: (a) channel
representation error due to the consideration of $\mathbf{X}_{\text{sp}}\in\mathbb{C}^{N\times N}$
instead of $\mathbf{X}\in\mathbb{C}^{N\times M}$, as described above,
for any UE channel, (b) channel representation error due to the algorithms
estimating a $D\times M$ (instead of $N\times M$) delay-angular
matrix representation for each UE (implying the ``missing'' $N-D$
columns are estimated as zeros), (c) channel estimation error due
to the channel representation error treated as an additional noise
term by the algorithm, and (d) channel estimation error due to the
increase of unknown parameters to be estimated. Note that these error
terms are controlled to a large extent by the design parameters $L_{1}$
and $L_{2}$. These should be selected to satisfy the two conflicting
requirements: reduce the sparse channel representation error (large
values for $L_{1}$ and $L_{2}$) and reduce the number of parameters
to be estimated (small values for $L_{1}$ and $L_{2}$). 

We numerically investigate the performance in the off-grid case and
the selection of $L_{1}$, $L_{2}$ in Sec. VI.

\section{Numerical Results}

For demonstrating the merits of the proposed, hierarchical-sparsity-based
framework for channel estimation, numerical examples are presented
in this section, demonstrating its effectiveness in achieving good
channel estimation accuracy with limited pilot overhead $N_{p}$. 

In all cases, an OFDM system with $N=1024$ subcarriers and a BS with
a ULA equipped with $M=256$ antennas are considered.  {Note that, for these system parameters, the channel transfer matrix of each UE consists of $MN = 262144$ elements, a huge number that imposes insurmountable computational challenges to conventional estimation approaches in addition to performance and overhead issues.} For all UEs,
the channel consists of $L$ paths and has a maximum delay spread
equal to $1/4$ of the useful OFDM symbol period. The channel
path gains for each UE were generated as i.i.d. zero mean, complex Gaussian variables
with a total power $\sum_{p=0}^{L-1}\mathbb{E}(|\rho_{p}|^{2})=1$,
resulting in an average  {received} power per subcarrier also equal to $1$ for all UEs (note
that the pilot signatures of the proposed design consist of unit-modulus symbols). The elements
of the noise matrix $\mathbf{Z}$ in (\ref{eq:observed_signal}) were
generated as i.i.d. zero mean, complex Gaussian random variables of
variance $1/\mathsf{SNR}$, where $\mathsf{SNR}$ denotes the average  {received}
signal-to-noise ratio per subcarrier. 

Towards minimizing the pilot
overhead,  {the F-S option will be considered throughout with $M_{p}=M$, i.e., all
antenna signals are used, unless stated otherwise}. In most examples, the HiIHT algorithm is
employed due to its simple implementation. The iterations of HiIHT and HiHTP terminate
when the estimated support between two consecutive iterations remains
the same or when ten iterations have been performed.

\subsection{The On-Grid Case}

\emph{Single User Case}: The single-UE case is considered first (i.e., $U=V=1$). The path
angles $\{\theta_{p}\}_{p=0}^{L-1}$ are generated independently and
uniformly over the angle sampling grid determined by $\mathbf{A}_{\theta}$,
however, no two paths are allowed to have the same angle, which is
a reasonable assumption for the asymptotic $M$ case. The path delays
$\{\tau_{p}\}_{p=0}^{L-1}$ are generated independently and uniformly
over the delay sampling grid determined by $\mathbf{A}_{\tau}$ with
$D=N/4=256$. Note that this channel model corresponds to an $(L,1,1)$-Hi-sparse
vector $\mathbf{x}\in \mathbb{C}^{M\cdot U \cdot D}$ in (\ref{eq:linear model}).

\begin{figure}[ptb]
\centering\includegraphics[width=1\columnwidth]{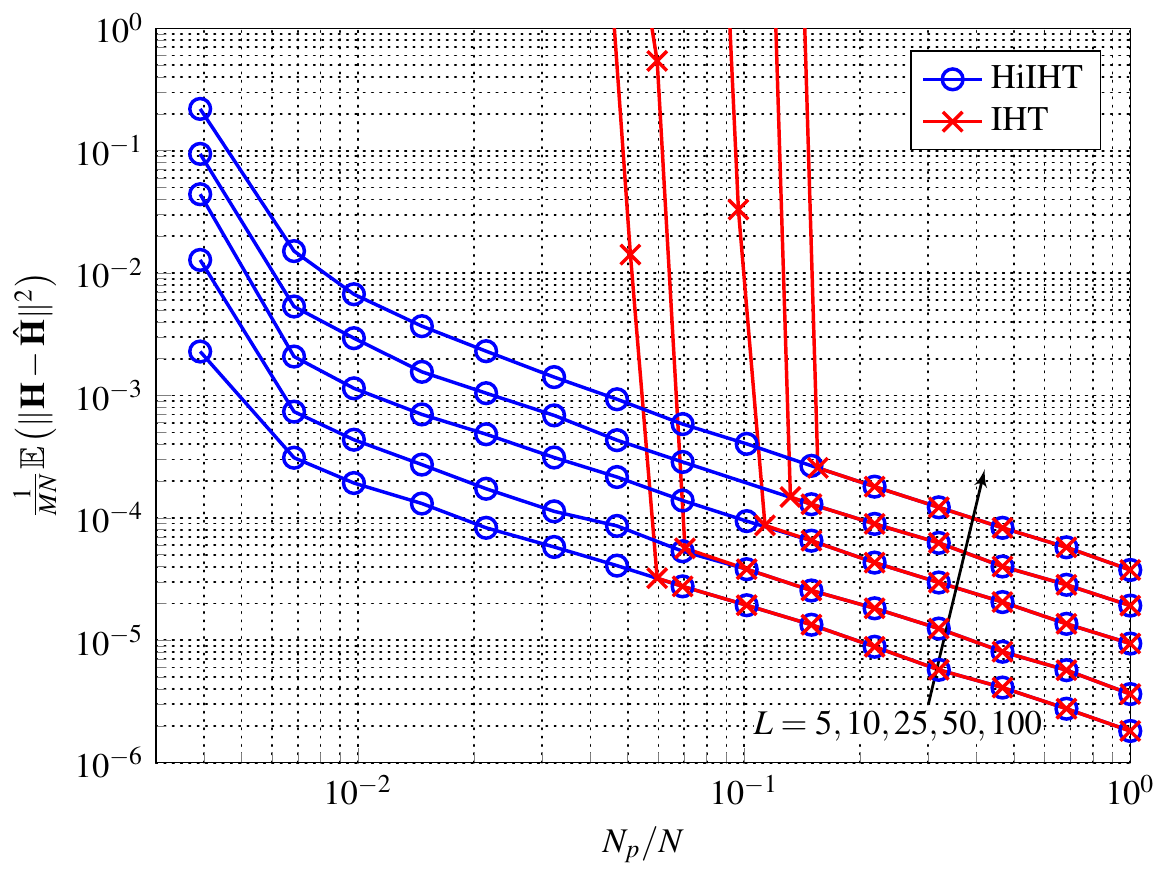}\caption{Single UE MSE of HiIHT and IHT estimators as a function of pilot overhead
(on-grid case, $N=1024,M=256,D=256,\mathrm{\mathsf{SNR}}=10\text{ dB}$).}
\label{fig:MSE_vs_O_tau}
\end{figure}

Figure \ref{fig:MSE_vs_O_tau} depicts the per-element mean squared
error (MSE) $\frac{1}{NM}\mathbb{E}(\|\mathbf{H}-\hat{\mathbf{H}}\|^{2})$
of the channel matrix estimate $\hat{\mathbf{H}}\triangleq\mathbf{A}_{\tau}\mathbf{\hat{X}}\mathbf{A}_{\theta}^{H}$,
where $\hat{\mathbf{X}}$ is the estimate of the delay-angular channel
representation provided by HiIHT, as a function of the normalized
pilot overhead $N_{p}/N$ and for various values of $L$ (assumed
known at the BS). The $\mathsf{SNR}$ was set equal to 10 dB.

It can be seen that HiIHT offers excellent estimation accuracy with
a very small pilot overhead. For example, a normalized pilot overhead
of around $10^{-2}$ is sufficient to achieve a MSE that is at least
one order of magnitude less than the noise variance level $1/\mathsf{SNR}=10^{-1}$,
which corresponds to the MSE achieved with $N_{p}=N$ and the naive 
channel estimate $\hat{\mathbf{H}}=\mathbf{Y}$. This pilot overhead
should be compared with conventional (non sparsity-exploiting) estimation
approaches which would require a normalized pilot overhead approximately
$D/N=0.25$ \cite{OFDM_chan_est_by_SVD} (see also discussion of Fig.
\ref{fig:MSE_vs_O_tau_offgrid}). As expected, the performance of
HiIHT degrades with increasing $L$ as the number of unknown parameters
increases.  {However, note that (a) this degradation is rather graceful,
i.e., the MSE remains bounded, and (b) the minimum required overhead to achieve a bounded MSE is independent of $L$, in line with the remarks made in the discussion of Theorem \ref{thm:required overhead}. Note also that
reasonable MSE is also achieved even with $L>N_{p}$. This reflects the advantage of observing multiple antennas and is in line with the flexibility in distributing overhead indicated by Theorem \ref{thm:required overhead}. Of course, in the single antenna case ($M=1$), reliable channel estimation can only be achieved with $N_p\geq L$.}

As a comparison, the performance of the standard IHT algorithm is
depicted in Fig. \ref{fig:MSE_vs_O_tau}. A ``phase transition''
phenomenon is clearly seen: a minimum pilot overhead is required in
order to achieve a reasonable MSE performance that is at least $6$
times greater than the one needed by HiIHT to achieve a MSE less than
$10^{-2}$. Also, this minimum overhead is increasing with $L$. This
clearly demonstrates the advantage of exploiting the hierarchical
channel sparsity in the channel estimation procedure, which allows
for reliable and robust performance in the small pilot overhead regime.
For sufficiently large training overhead, the performances of IHT and
HiIHT are the same, implying that knowledge of the sparsity structure
plays no role in this regime. This is in line to the well-known fact
from estimation theory that \emph{a priori} information (in this case, hierarchical structure of sparsity) becomes irrelevant once
sufficiently many observations have been obtained.

\begin{figure}[ptb]
\centering\includegraphics[width=1\columnwidth]{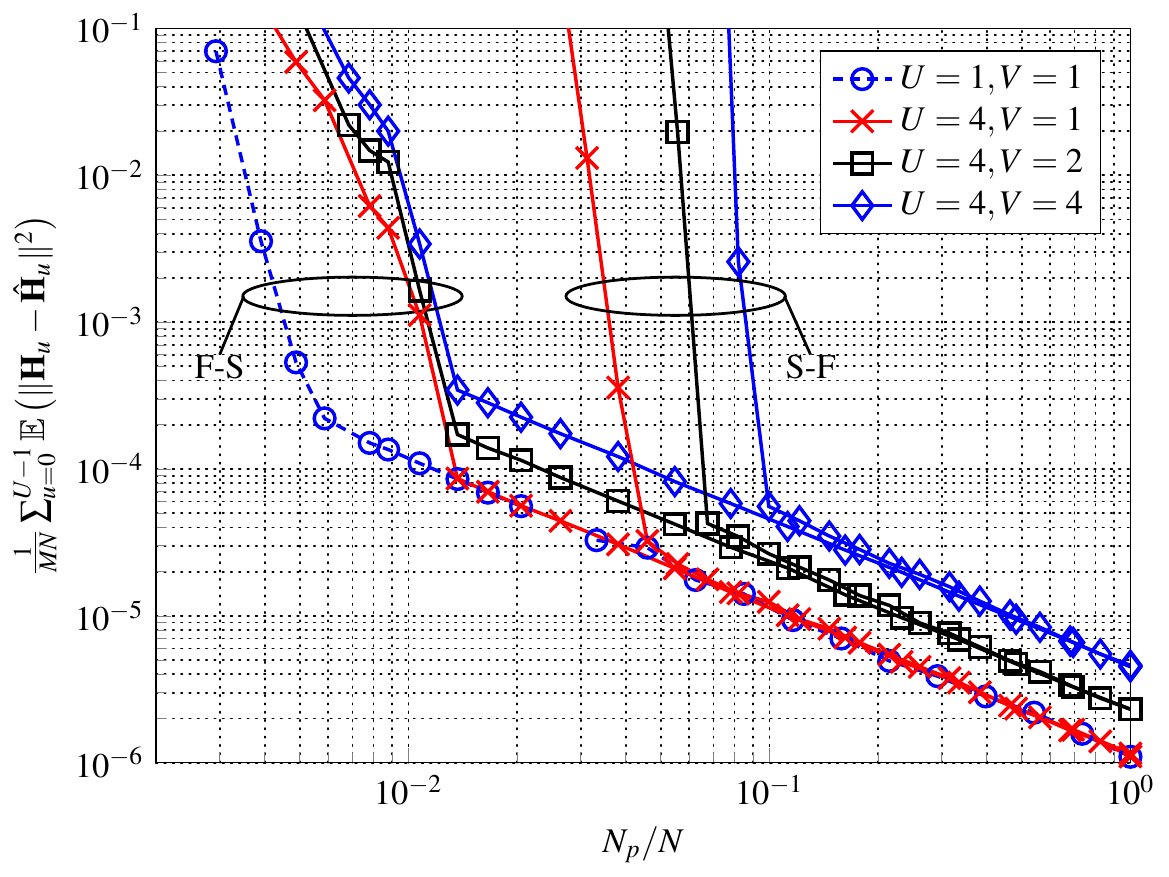}\caption{Multiuser MSE of HiIHT estimator as a function of pilot overhead (on-grid
case, $N=1024,M=256,D=256,L=3,\mathrm{\mathsf{SNR}}=10\text{ dB}$).}
\label{fig:MSE_vs_O_tau_MC}
\end{figure}

\emph{Multiuser Case}: The multiuser case is considered next. Note that for the scenario
with $D=N/4$ considered here, up to $U=4$ UEs can be supported per
UE group by the pilot sequence assignment scheme of Sec. IV. With
 $\mathsf{SNR}=10$ dB and UE channels
with $L=3$ paths generated independently as described in the single
UE case example, Fig. \ref{fig:MSE_vs_O_tau_MC} demonstrates the
total MSE, defined as $\frac{1}{MN}\sum_{u=0}^{U-1}\mathbb{E}(\|\mathbf{H}_{u}-\hat{\mathbf{H}}_{u}\|^{2})$,
for various number  of randomly and uniformly selected  active UEs  $V\leq U$. Note that, in
the MSE formula, $\mathbf{H}_{u}$ is equal to zero if the UE is not active. 

When only one UE is active, i.e., $V=1$, a slightly larger pilot
overhead compared to the single UE case ($U=V=1)$ is required to
achieve the same MSE performance. This overhead cost can be attributed
to the uncertainty at the BS of who the actual active UE is. By increasing
$V$, a degradation of MSE performance is observed that is proportional
to $V$ due to the corresponding increase of unknown parameters to
be estimated.  {However, the minimum required overhead to achieve a bounded channel estimation error is independent of $V$, as  guaranteed by Theorem \ref{thm:required overhead}}. 

 {Figure \ref{fig:MSE_vs_O_tau_MC} also depicts the MSE performance under the S-F option. For this case, the channel vector $\mathbf{x}$ in (\ref{eq:linear model}) was generated as an $(V,L,1)$-Hi-sparce vector in $\mathbb{C}^{U\cdot D \cdot M}$ with path gains having the same statistics as in the channel model considered under the F-S option. It can be seen that this approach (a) requires increased training overhead to achieve reliable channel estimation and (b) the minimum overhead increases with $V$. Both these observations are consistent with Theorem \ref{thm:required overhead}.}

\begin{figure}[ptb]
\centering\includegraphics[width=1\columnwidth]{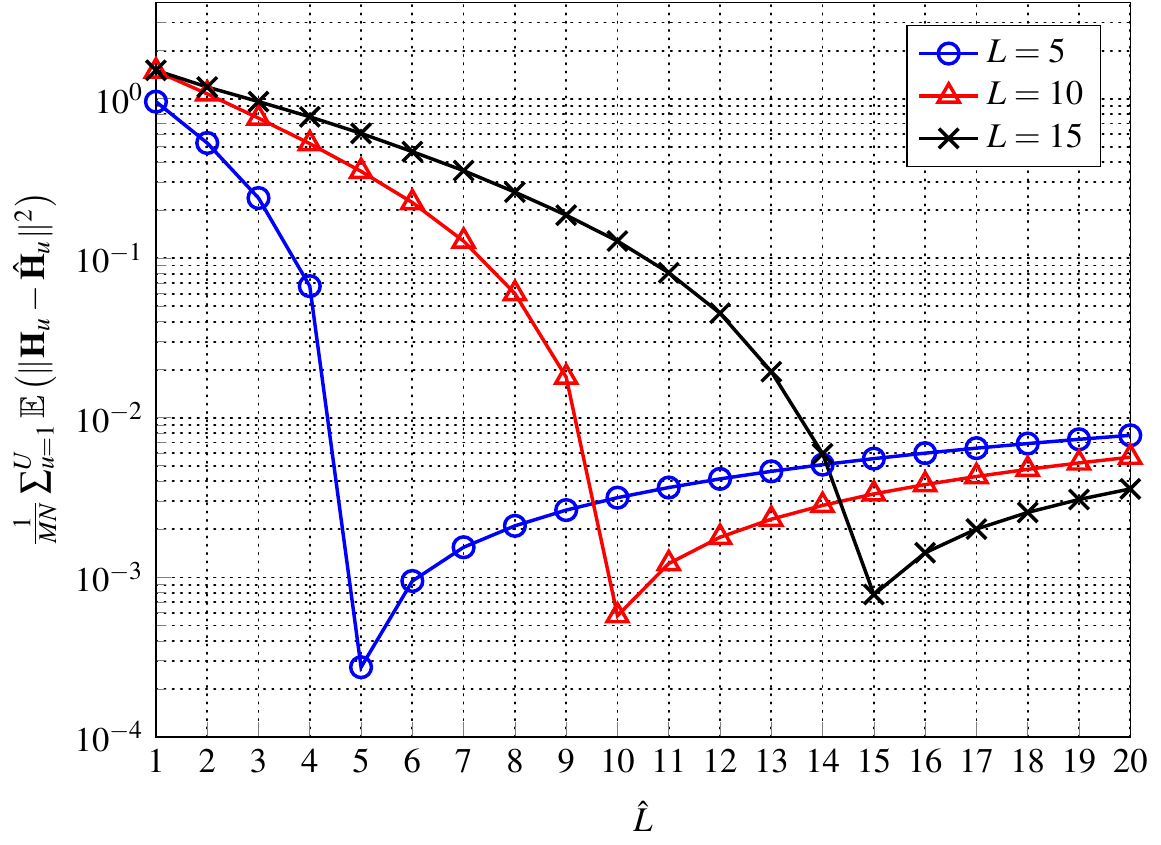}\caption{ {Performance of HiIHT under mismatched $L$ ($N=1024,M=256,D=256, N_p=15, U=4,V=2,\mathsf{SNR}=10\text{ dB}$).}}
\label{fig:mismatched_L}
\end{figure}

 {\emph{Unknown $L$}:  Both the analysis and the previous results assume knowledge of the number of channel paths $L$.  Figure \ref{fig:mismatched_L}  shows the performance of HiIHT assuming $\hat{L}$ number of paths instead of $L$. A case with $U=4, V=2$, and $N_p=15$ pilot subcarriers is considered with the rest of the system parameters same as above. As expected, there is degradation in MSE when $\hat{L}\neq L$. This degradation is much more prominent when $\hat{L}< L$, whereas $\hat{L}> L$ results in a moderate degradation. This suggests that setting $\hat{L}$ as an upper bound (worst case) value could be a practical approach when $L$ is unknown. Another approach is to modify HiIHT/HiHTP so as it also provides an estimate of $L$, as described in, e.g.,  \cite{unknown_L1, unknown_L2}.}

% However, we note that the HiIHT/HiHTP algorithms can naturally be modified so as an on-the-fly estimate of $L$ is also provided. Note from the description of the algorithms that the error $\|\mathbf{y} - \mathbf{A} \hat{\mathbf{x}}^{(i)}\|^2$ is computed for each iteration $i$ of the algorithm and this error will be equal to $\|\mathbf{z}\|^2$ if the channel estimate is perfect. By the law of large numbers, the latter value is approximately equal to $\mathbb{E}(\|\mathbf{z}\|^2)=1/(N_p M_p \mathsf{SNR})$. Therefore, assuming the training overhead is sufficiently large to guarantee reliable channel estimation (under perfect knowledge of $L$), one approach is to run the algorithm for multiple values of $\hat{L}$ and choose the estimate corresponding to the smallest $\hat{L}$ achieving an error $\|\mathbf{y} - \mathbf{A} \hat{\mathbf{x}}^{(i^*)}\|^2$ at the final iteration $i^*$ that is close to $1/(N_p M_p \mathsf{SNR})$. Figure \ref{fig:mismatched_L}  shows the value of the error $\|\mathbf{y} - \mathbf{A} \hat{\mathbf{x}}^{(i^*)}\|^2$ averaged over 10 independent channel realizations. It can be seen that increasing $\hat{L}$ decreases this error until $\hat{L}= L$. For greater values of $\hat{L}$, the error decrease is very small as the algorithm has already captured the effect of the user channels on $\mathbf{y}$ in the strongest $VL$ elements of $\hat{\mathbf{x}}^{(i^*)}$ with the remaining $V(\hat{L}-L)$ capture spurious, minor energy contributions of the wideband noise.}

\begin{figure}[ptb]
\centering\includegraphics[width=1\columnwidth]{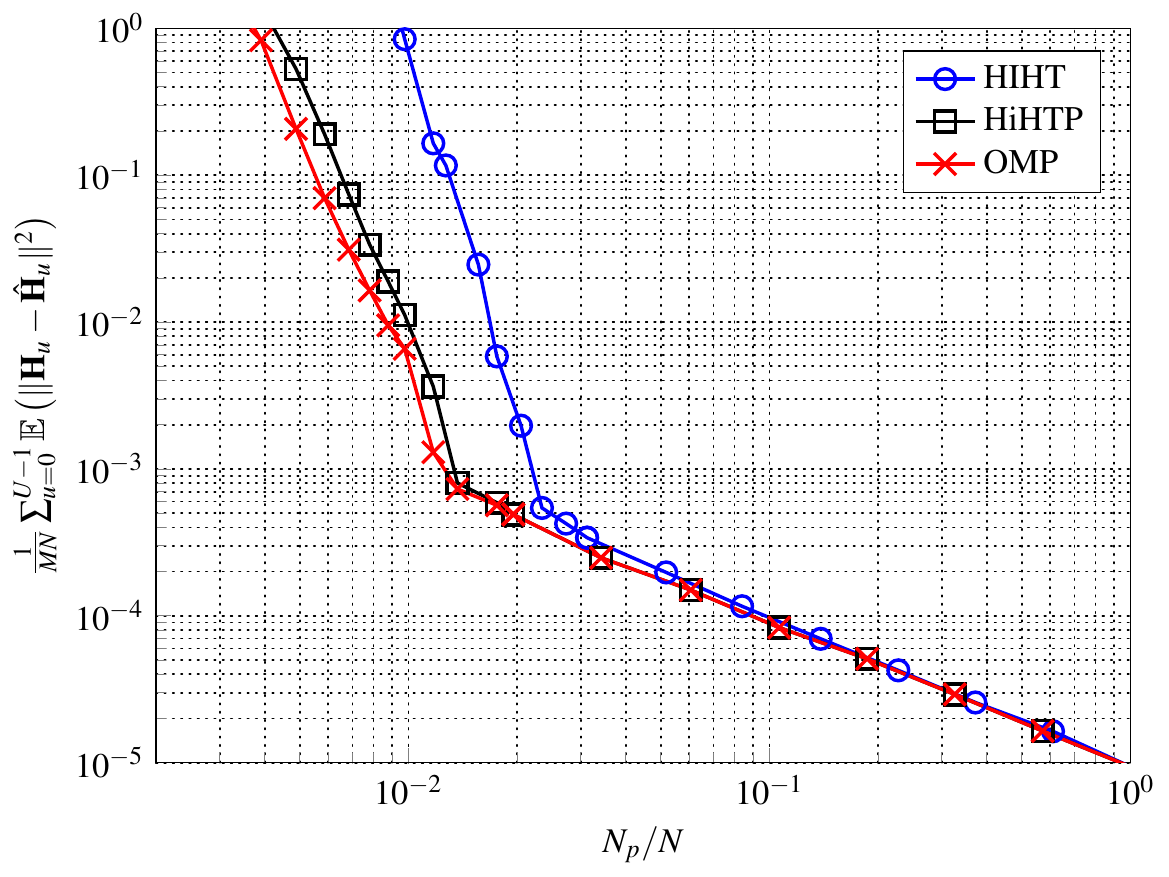}\caption{ {Performance comparison of HiHTP, HiIHT, and OMP ($N=1024,M=256,D=256, L=3, M_p=64, U=4,V=2,\mathsf{SNR}=10\text{ dB}$).}}
\label{fig:OMP}
\end{figure}

 {\emph{Comparison with Orthogonal Matching Pursuit (OMP)}: In this example, we compare the proposed HiIHT/HiHTP algorithms with the commonly employed OMP algorithm \cite{MathIntroToCS}, which  forms the basis for many previously proposed massive MIMO channel estimation schemes \cite{Heath JSAC 2017, Heath how many measurements}. OMP is a greedy, iterative algorithm of roughly the same complexity as HiHTP. It ignores any structural properties of  sparsity, i.e.,  treats the unknown vector in (\ref{eq:linear model}) as $VL$-sparse, instead of $(VL,1,1)$-Hi-sparse (the F-S option is considered). Simulations (not shown here) with $M_p=M$ showed that OMP achieved the exact same performance as HiIHT and HiHTP. Towards identifying performane differences, we considered a case with $M_p=M/4=64$, with the remaining system parameters same as above and the results are shown in Fig. \ref{fig:OMP}. It can be seen that, in this scenario, OMP is a competitive alternative of HiHTP, whereas HiIHT performs slightly worse but is significantly less complex. The good performance of OMP in estimating hierarchically sparse vectors, even though not explicitly taking this property into account,  was previously identified analyzed in \cite{OMP_hisparse}. This close correspondence of OMP with HiHTP/HiIHT suggests that the analytical results in this paper may have broader applicability than the HiHTP/HiIHT algorithms. We leave this topic for future investigation.}

\subsection{The Off-Grid case}

\begin{figure}[ptb]
\centering\includegraphics[width=1\columnwidth]{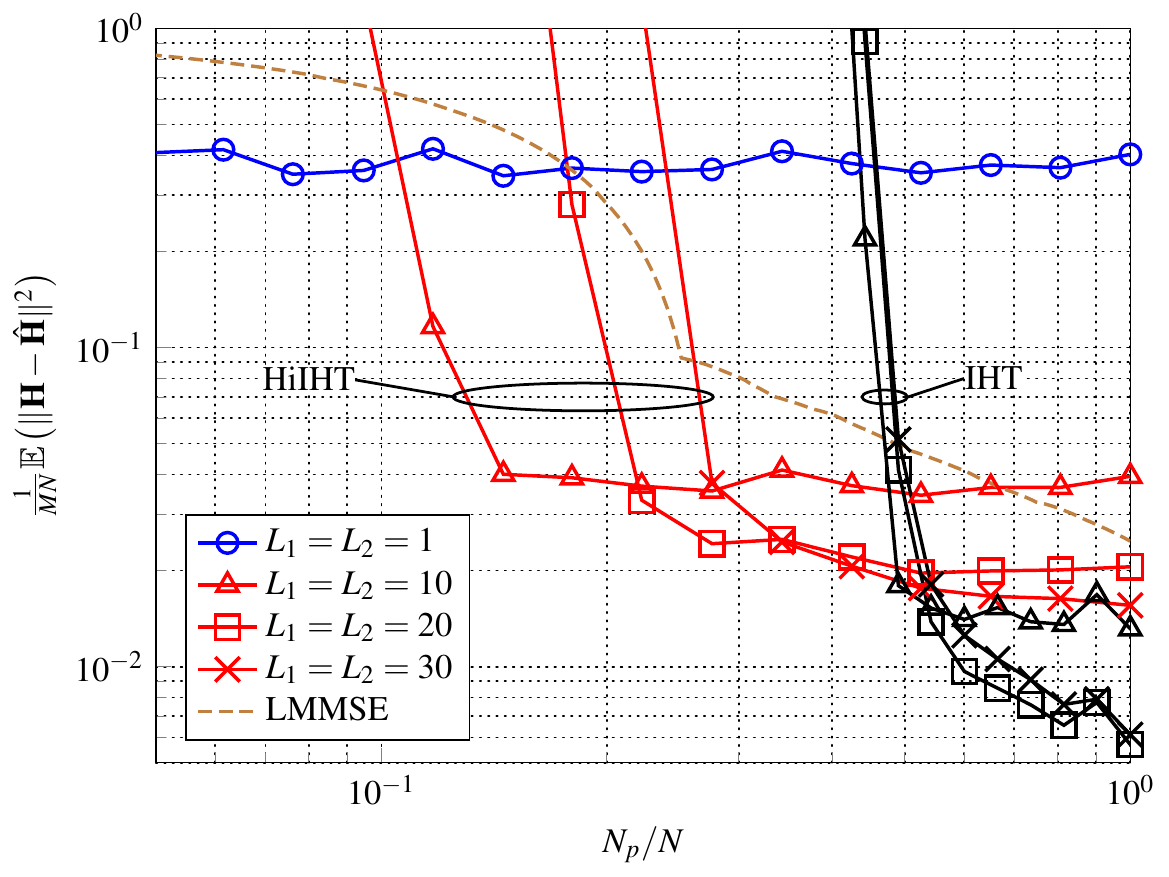}\caption{Single UE MSE of HiHTP, HTP and LMMSE estimators as a function of
pilot overhead (off-grid case, $N=1024,M=256,D=256,\mathrm{SNR}=10\text{ dB}$).}
\label{fig:MSE_vs_O_tau_offgrid}
\end{figure}

Figure \ref{fig:MSE_vs_O_tau_offgrid} demonstrates the single UE
MSE performance for the off-grid case, where the channel is generated
as described in the on-grid case with $L=3$, however, with the paths
angles and delays uniformly and independently distributed over the
continuous domains $[0,1)$ and $[0,T_{S}/4)$, respectively. The
$\mathsf{SNR}$ was set to $10$ dB.

As can be seen, performance of HiIHT strongly depends on the choice
of $L_1$ and $L_2$. Small values of these parameters
result in the estimation of a small number of unknown parameters by
the channel estimator, however, with the cost of a large sparse channel
approximation error. As can be seen, the optimal values of $L_1$,
$L_2$ are proportional to the pilot overhead, which is expected
as increasing the latter allows for the reliable estimation of more
parameters. In any case, the basis mismatch effect results in a great
performance degradation compared to the idealized, on-grid examples
presented above. However, a MSE of almost an order of magnitude less
that the noise level is achievable, rendering the effect of the channel estimation error negligible at the decoding stage.

For comparison, the performance of the conventional, linear minimum
mean squares estimator (LMMSE) estimator with equally-spaced pilot
subcarriers is depicted in Fig. \ref{fig:MSE_vs_O_tau_offgrid}. The
LMMMSE estimator utilizes only information about the correlation function
$\mathbb{E}([\mathbf{H}]_{n,m}[\mathbf{H}]_{n^{\prime},m^{\prime}}^{*})$
for $n,n^{\prime}\in[N]$, $m,m^{\prime}\in[M]$. The latter can be
obtained by a straightforward generalization of the approach shown
in \cite{OFDM_chan_est_by_SVD} for the single receive antenna case.
It can be seen that the LMMSE estimator performs very poorly, requiring
at least $N_{p}/N\approx0.25$ in order to achieve an MSE that is
equal to the noise level. This is due to the correlation function
not capturing the sparsity properties of the channel. Figure \ref{fig:MSE_vs_O_tau_offgrid}
also shows the performance of the standard IHT algorithm operating
assuming $LK_{\tau}K_{\theta}$ on-grid paths, i.e., the same number
of on-grid paths considered by HiIHT. It can be seen that IHT provides
a reasonable performance only for large pilot overhead (greater than
$0.4$). In that regime, it actually provides a better MSE than HiIHT
suggesting that the hierarchical sparsity structure assumed by the
HiIHT is not accurate, resulting in an additional error term introduced
in the estimate due to this mismatch. However, even though not accurate
the assumption of hierarchical sparsity is beneficial in the small
pilot overhead regime.

\section{Conclusion}
The problem of channel estimation for multiuser wideband massive MIMO via a compressive sensing approach was investigated. Under the assumption of on-grid channel param- eters, a problem reformulation that highlights the hierarchical sparsity property of the wireless channel was considered. This property was taken into account for the design of low- complexity channel estimation algorithms. Using the HiRIP analysis framework, rigorous performance guarantees for these algorithms were obtained that, in turn, provide design rules for UE pilot signature design and selection of pilot subcarriers. A characterization of the sufficient pilot overhead required to achieve reliable channel estimation was provided, revealing that in the massive MIMO regime, the number of subcarriers is independent from the number of active UEs and channel paths per UE. These observations were also verified numerically, with the proposed algorithm showing significant performance gain over conventional CS approaches of similar complexity. Application of the algorithms in a multiple measurements and off-grid channel parameter setting was discussed. For the later case, which is valid in the finite antenna and bandwidth regime, even though there exists an error due to model mismatch, performance of proposed algorithms is still significantly better from conventional CS as well as the standard LMMSE approach.

\appendices{}

\section{\label{sec:proof of Eq 14} Proof of (\ref{eq:HIRIP<=00003DRIP})}
 {Let $\{(N_i,s_i)\}_{i=1}^{\ell}$ be a set of $\ell$ tuples of integers such that $N_i \geq s_i\geq 1$ for all $i$. Denote  $\mathcal{S}_{(s_1,s_2,\ldots,s_\ell)}\subseteq \mathbb{C}^{N_1\cdot N_2 \cdots N_\ell}$ the set of all $(s_1,s_2,\ldots,s_\ell)$-Hi-sparse vectors in $\mathbb{C}^{N_1\cdot N2 \cdots N_\ell}$, and $\mathcal{S}_{s_1s_2\cdots s_\ell}\subseteq \mathbb{C}^{N_1 N_2 \cdots N_\ell}$ the set of all $s_1s_2\cdots s_\ell$-sparse vectors in $\mathbb{C}^{N_1 N2 \cdots N_\ell}$. Note that $\mathcal{S}_{(s_1s_2\cdots s_\ell)} \subseteq \mathcal{S}_{s_1 s_2 \cdots s_\ell}$.  For an arbitrary matrix $\tilde{\mathbf{A}} \in \mathbb{C}^{N_0 \times (N_1 N_2 \cdots N_\ell)}$, $N_0\geq 1$, it follows from the definition of the HiRIP and RIP constants that
\begin{align*}
    \delta_{(s_1, s_2, \ldots, s_\ell)}(\tilde{\mathbf{A}}) &= \underset{\mathbf{x} \in \mathcal{S}_{(s_1,s_2,\ldots,s_\ell)}}{\max} \frac{\left| \| \tilde{\mathbf{A}} \mathbf{x} \|^2 - \|\mathbf{x}\|^2 \right|}{\|\mathbf{x}\|^2}\\ &\leq  \underset{\mathbf{x} \in \mathcal{S}_{s_1 s_2 \cdots s_\ell}}{\max} \frac{\left| \| \tilde{\mathbf{A}} \mathbf{x} \|^2 - \|\mathbf{x}\|^2 \right|}{\|\mathbf{x}\|^2}\\ &= \delta_{s_1 s_2  \cdots, s_\ell}(\tilde{\mathbf{A}}).
\end{align*}}

\section{\label{sec:proof of required overhead}Proof of Theorem \ref{thm:required overhead}}

\noindent Let $\tilde{\mathbf{x}}\neq\mathbf{0}\in\mathbb{C}^{UD}$
denote the $s$-sparse vector for which $\left|\|\bar{\mathbf{A}}_{\tau}\tilde{\mathbf{x}}\|^{2}-\|\tilde{\mathbf{x}}\|^{2}\right|=\delta_{s}(\bar{\mathbf{A}}_{\tau})\|\tilde{\mathbf{x}}\|^{2}$,
where $\delta_{s}(\bar{\mathbf{A}}_{\tau})$ is the $s$-RIP constant
of matrix $\bar{\mathbf{A}}_{\tau}$ given in (\ref{eq:A_tau_with_proposed_design}).
Let $\tilde{\mathbf{x}}_{\text{ext}}\triangleq[\tilde{\mathbf{x}}^{T},\mathbf{0}^{T}]^{T}\in\mathbb{C}^{N}$
denote its zero padded extension that is also $s$-sparse. Consider
$\bar{\mathbf{A}}_{\tau,\text{ext}}\triangleq(1/\sqrt{N_{p}})\mathbf{P}_{\mathcal{N}_{p}}\text{diag}(\mathbf{c})\mathbf{F}_{N,N}$.
It holds
\begin{align*}
\delta_{s}(\bar{\mathbf{A}}_{\tau})\|\tilde{\mathbf{x}}\|^{2}  &=\left|\|\bar{\mathbf{A}}_{\tau}\tilde{\mathbf{x}}\|^{2}-\|\tilde{\mathbf{x}}\|^{2}\right|\\
 &=\left|\|\bar{\mathbf{A}}_{\tau,\text{ext}}\tilde{\mathbf{x}}_{\text{ext}}\|^{2}-\|\tilde{\mathbf{x}}_{\text{ext}}\|^{2}\right|\\
 & \leq\underset{s\text{-sparse }\mathbf{p}\in\mathbb{C}^{N},\|\mathbf{p}\|=\|\tilde{\mathbf{x}}\|}{\max}\left|\|\bar{\mathbf{A}}_{\tau,\text{ext}}\mathbf{p}\|^{2}-\|\mathbf{p}\|^{2}\right|\\ &\leq\delta_{s}(\bar{\mathbf{A}}_{\tau,\text{ext}})\|\tilde{\mathbf{x}}\|^{2},
\end{align*}
where $\delta_{s}(\bar{\mathbf{A}}_{\tau,\text{ext}})$ is the $s$-RIP
constant of matrix $\bar{\mathbf{A}}_{\tau,\text{ext}}$, resulting
in
\begin{equation}
\delta_{s}(\bar{\mathbf{A}}_{\tau})\leq\delta_{s}(\bar{\mathbf{A}}_{\tau,\text{ext}}),\text{ for all }s\leq UD.\label{eq:bound of RIP by extended matrix RIP}
\end{equation}

\noindent Now consider the F-S option, i.e., with $\mathbf{A}=\bar{\mathbf{A}}_{\theta}^{*}\otimes\bar{\mathbf{A}}_{\tau}$
acting on $\mathbf{x}\in\mathbb{C}^{M\cdot U\cdot D}$. It holds
\begin{align*}
 &\delta_{(3VL,3K_{V},3K_{L})}(\mathbf{A})\\
 \overset{(a)}{\leq} & \left(1+\delta_{3VL}\left(\bar{\mathbf{A}}_{\theta}^{*}\right)\right)\left(1+\delta_{9K_{V}K_{L}}\left(\bar{\mathbf{A}}_{\tau}\right)\right)-1\\
 \overset{(b)}{\leq} & \left(1+\delta_{3VL}\left(\bar{\mathbf{A}}_{\theta}^{*}\right)\right)\left(1+\delta_{9K_{V}K_{L}}\left(\bar{\mathbf{A}}_{\tau,\text{ext}}\right)\right)-1,
\end{align*}

\noindent where $(a)$ follows from Theorem \ref{thm:HiRIP bound}
and ($b)$ from (\ref{eq:bound of RIP by extended matrix RIP}). Therefore,
a sufficient condition for (\ref{eq:HiRIP requirement}) to hold is
\begin{align}
1/\sqrt{3} & >\left(1+\delta_{3VL}\left(\bar{\mathbf{A}}_{\theta}^{*}\right)\right)\left(1+\delta_{9K_{V}K_{L}}\left(\bar{\mathbf{A}}_{\tau,\text{ext}}\right)\right)-1.\label{eq:app_eq_1}
\end{align}

\noindent For any $\delta_{\tau}\in(0,1)$ and $\delta_{\theta}\in(0,1)$,
set $N_{p}$ and $M_{p}$ as in (\ref{eq:O_tau min}) and (\ref{eq:O_theta min}),
respectively. Noting that $\bar{\mathbf{A}}_{\theta}^{*}$ and $\bar{\mathbf{A}}_{\tau,\text{ext}}$
are obtained by random sampling of the rows of orthonormal matrices,
it follows from \cite[Theorem 12.31]{MathIntroToCS} that $\delta_{9K_{V}K_{L}}\left(\bar{\mathbf{A}}_{\tau,\text{ext}}\right)<\delta_{\tau}$
and $\delta_{3VL}\left(\bar{\mathbf{A}}_{\theta}^{*}\right)<\delta_{\theta}$
with probability larger $1-N^{-\log^{3}(N)}$ and $1-M^{-\log^{3}(M)}$,
respectively, which results in an upper bound for the right hand side
expression of (\ref{eq:app_eq_1}) equal to $\delta_{\tau}+\delta_{\theta}+\delta_{\tau}\delta_{\theta}$.
Selecting values for $\delta_{\theta}$ and $\delta_{\tau}$ such
that this upper bound is less than $1/\sqrt{3}$ immediately implies
(\ref{eq:HiRIP requirement}). The proof for the S-F case follows
the exact same steps.

\section{\label{sec:Proof_of_sparsification_if_off_grid_case}Proof of Theorem
\ref{thm:sparsification_in_off_grid_case}}

For an arbitrary channel transfer matrix $\mathbf{H}\in\mathbb{C}^{N\times M}$
assuming $D=N$, it follows from (\ref{eq:channelmatrix}) and (\ref{eq:delay-angular_representation_off_grid})
that its delay-angular representation equals
$\mathbf{X} =\sum_{p=0}^{L-1}\rho_{p}\mathbf{u}_{N}(\tilde{\tau}_{p})\mathbf{u}_{M}^{H}(\theta_{p})$, where $\tilde{\tau}_{p}\triangleq\tau_{p}/T_{s}\in[0,1]$
is the normalized delay of the $p$-th path and $\mathbf{u}_{K}:[0,1]\rightarrow\mathbb{C}^{K}$
with \cite{ChenYang206_TWC}
\[
\left[\mathbf{u}_{K}(\omega)\right]_{k}\triangleq\frac{\sin\left(\pi K(\omega-k/K)\right)}{K\sin\left(\pi(\omega-k/K)\right)}e^{-j\pi\left(K-1\right)(\omega-k/K)},k\in[K].
\]

We consider a sparse approximation of $\mathbf{X}$ given by $\mathbf{X}_{\text{sp}}=\sum_{p=0}^{L-1}\rho_{p}\mathbf{u}_{N,\text{sp}}(\tilde{\tau}_{p};L_{1})\mathbf{u}_{M,\text{sp}}^{H}(\theta_{p};L_{2})$, where $\mathbf{u}_{K,\text{sp}}(\omega;J)\in\mathbb{C}^{K}$
is a $(2J+1)$-sparse vector obtained by retaining the $(2J+1)$ largest
modulus elements of $\mathbf{u}_{K}(\omega)$ and the rest elements
set equal to zero. Note that with this construction, $\mathbf{X}_{\text{sp}}$
can have at most $L(2L_{1}+1)(2L_{2}+1)$ non-zero elements. In order
to investigate the sparse approximation error, we first focus on quantifying
the error $\|\mathbf{u}_{M,\text{sp}}(\theta;L_{2})-\mathbf{u}_{M}(\theta)\|$for
any $\theta\in[0,1]$. It is easy to see that the non-zero elements
of $\mathbf{u}_{M,\text{sp}}(\theta;L_{2})$ are consecutive in a
wrap-around sense (i.e., the element indices $0$ and $M-1$ are assumed
consecutive). By symmetry, it is sufficient to consider the error
for some value of $\theta\in[0,\frac{1}{2M}]$. In this case, the
set of non-zero elements of $\mathbf{u}_{M,\text{sp}}(\theta;L_{2})$
is $\mathcal{A}=\{0,1,\ldots L_{2}\}\cup\{M-1-L_{2},M-L_{2},\ldots,M-1\}$
and it holds
\begin{align}
 &\|\mathbf{u}_{M,\text{sp}}(\theta;L_{2})-\mathbf{u}_{M}(\theta)\|^{2}\\
= &\sum_{m\in[M]\setminus\mathcal{A}}\left|\left[\mathbf{u}_{M}(\theta)\right]_{m}\right|^{2}\nonumber \\
\leq  &\sum_{m\in[M]\setminus[L_{2}+1]}\left|\left[\mathbf{u}_{M}(\theta)\right]_{m}\right|^{2}\nonumber \\
= & \frac{1}{M^{2}}\sum_{m\in[M]\setminus[L_{2}+1]}\frac{\sin^{2}(\pi M(\theta-m/M))}{\sin^{2}(\pi(\theta-m/M))}\label{eq:ap2_eq}\\
\overset{(a)}{\leq}  &\frac{1}{M^{2}}\sum_{m\in[M]\setminus[L_{2}+1]}\frac{(M+1)^{2}}{(M-1)^{2}4(\theta-m/M)^{2}}\label{eq:ap2_eq2}\\
\overset{(b)}{\leq}  &\frac{(M+1)^{2}}{(M-1)^{2}}\sum_{m\in[M]\setminus[L_{2}+1]}\frac{1}{(2m-1)^{2}}\nonumber \\
\leq  &\frac{(M+1)^{2}}{(M-1)^{2}}\int_{L_{2}+1}^{M-1}\frac{1}{(2x-1)^{2}}dx\nonumber \\
=  &\frac{(M+1)^{2}}{(M-1)^{2}}\frac{1}{2}\left(\frac{1}{2L_{2}+1}-\frac{1}{2M-3}\right)\nonumber \\
\leq &\frac{1}{L_{2}},\nonumber 
\end{align}

\noindent where $(a)$ follows by trivially upper bounding the numerator
of the summand in (\ref{eq:ap2_eq}) by $1$ and by lower bounding
the denominator according to the inequality $\sin^{2}(\pi x)\geq4x^{2}(M-1)/(M+1),$
which holds for all $|x|\leq\pi/2+1/(2M)$, $(b)$ follows by minimizing
the term $(\theta-m/M)^{2}$ in the summand of (\ref{eq:ap2_eq2})
w.r.t. $\theta\in[0,1/(2M)]$ and the last inequality holds for $M\geq3$,
which can be safely assumed to hold in massive MIMO applications.
Note that the obtained bound holds for any $\theta\in[0,1]$. In the
exact same fashion, it can be proved that $\|\mathbf{u}_{N,\text{sp}}(\tilde{\tau};L_{1})-\mathbf{u}_{N}(\tilde{\tau})\|^{2}\leq1/L_{1}$
for any $\tilde{\tau}\in[0,1]$. Now, for any $\tilde{\tau}$, $\theta$,
and dropping, for simplicity, the arguments from the notation of $\mathbf{u}_{N}(\tilde{\tau})$,
$\mathbf{u}_{M}(\theta)$, $\mathbf{u}_{N,\text{sp}}(\tilde{\tau};L_{1})$,
$\mathbf{u}_{M,\text{sp}}(\theta;L_{2})$ it holds
\begin{align}
 &\left\Vert \mathbf{u}_{N}\mathbf{u}_{M}^{H}-\mathbf{u}_{N,\text{sp}}\mathbf{u}_{M,\text{sp}}^{H}\right\Vert\\ 
\overset{(a)}{\leq} &\left\Vert \mathbf{u}_{N}\mathbf{u}_{M}^{H}-\mathbf{u}_{N}\mathbf{u}_{M,\text{sp}}^{H}\right\Vert +\left\Vert \mathbf{u}_{N}\mathbf{u}_{M,\text{sp}}^{H}-\mathbf{u}_{N,\text{sp}}\mathbf{u}_{M,\text{sp}}^{H}\right\Vert \nonumber \\
\overset{(b)}{\leq}  &\|\mathbf{u}_{N}\|\left\Vert \mathbf{u}_{M}^{H}-\mathbf{u}_{M,\text{sp}}^{H}\right\Vert +\|\mathbf{u}_{M,\text{sp}}\|\left\Vert \mathbf{u}_{N}-\mathbf{u}_{N,\text{sp}}\right\Vert \nonumber \\
\overset{(c)}{\leq}  &\|\mathbf{u}_{N}\|\left\Vert \mathbf{u}_{M}^{H}-\mathbf{u}_{M,\text{sp}}^{H}\right\Vert +\|\mathbf{u}_{M}\|\left\Vert \mathbf{u}_{N}-\mathbf{u}_{N,\text{sp}}\right\Vert\\
\overset{(d)}{\leq}  &\frac{1}{\sqrt{L_{1}}}+\frac{1}{\sqrt{L_{2}}},\label{eq:app.eq}
\end{align}

\noindent where $(a)$ follows from the triangle inequality, $(b)$
from the Cachy-Schwarz inequality, $(c)$ by noting that $\|\mathbf{u}_{M,\text{sp}}\|\le\|\mathbf{u}_{M}\|$
and $(d)$ by noting that $\|\mathbf{u}_{M}\|=\|\mathbf{u}_{N}\|=1$
and using the bounds for $\left\Vert \mathbf{u}_{N}-\mathbf{u}_{N,\text{sp}}\right\Vert $
and $\left\Vert \mathbf{u}_{M}-\mathbf{u}_{M,\text{sp}}\right\Vert $
obtained above. The sparse approximation error of $\mathbf{X}_{\text{sp}}$
can now be obtained as
\begin{align*}
  &\|\mathbf{X}_{\text{sp}}-\mathbf{X}\|\\
=  &\left\Vert \sum_{p=0}^{L-1}\rho_{p}\left[\mathbf{u}_{N}\left(\tilde{\tau}\right)\mathbf{u}_{M}^{H}(\theta)-\mathbf{u}_{N,\text{sp}}\left(\tilde{\tau};L_{1}\right)\mathbf{u}_{M,\text{sp}}^{H}(\theta;L_{2})\right]\right\Vert \\
 \leq  &\sum_{p=0}^{L-1}\left|\rho_{p}\right|\left\Vert \mathbf{u}_{N}\left(\tilde{\tau}\right)\mathbf{u}_{M}^{H}(\theta)-\mathbf{u}_{N,\text{sp}}\left(\tilde{\tau};L_{1}\right)\mathbf{u}_{M,\text{sp}}^{H}(\theta;L_{2})\right\Vert .
\end{align*}

\noindent Applying (\ref{eq:app.eq}) results in (\ref{eq:off_grid_sparsification_error_bound}).

\end{document}